\documentclass[aps,prb,twocolumn,floatfix,longbibliography,superscriptaddress]{revtex4-2}

\usepackage{graphicx}
\usepackage{amsmath}
\usepackage{amssymb}
\usepackage{braket}
\usepackage{bbold}
\usepackage{hyperref}
\usepackage{subfigure}
\usepackage[usenames,dvipsnames]{color}
\usepackage{graphicx}
\usepackage{subfigure}
\usepackage[mathscr]{euscript}
\hypersetup{colorlinks,allcolors=black}

\begin{document}
\title{Interfacial Line Energy of a Topological Phase}

\author{Saikat Mondal}
\email{msaikat@iitk.ac.in}
\affiliation{Department of Physics, Indian Institute of Technology Kanpur, Kalyanpur, Uttar Pradesh 208016, India}

\author{Adhip Agarwala}
\email{adhip@iitk.ac.in}
\affiliation{Department of Physics, Indian Institute of Technology Kanpur, Kalyanpur, Uttar Pradesh 208016, India}

\begin{abstract}
In interacting topological systems, Landau-like order parameters interplay with the band topology of fermions. The physics of domain formation in such systems can get significantly altered due to the presence of topological fermions. In this work we show that coupling a topological fermionic field to a scalar field can drastically modify the nucleation processes. We find that existence of non-trivial fermionic boundary modes on the nucleating droplets of the scalar field leads to substantial quantum corrections to the interface energy. This leads to an increase in the size of the critical nucleus beyond which unrestricted droplet growth happens. To illustrate the phenomena we devise a minimal model of fermions in a Chern insulating system coupled to a classical Ising field in two spatial dimensions. Using a combination of analytic and numerical methods we conclusively demonstrate that topological phases have a characteristic quantum correction to the interfacial line energy. Apart from material systems, our work opens up a host of questions regarding the impact of fermionic topological terms on classical phase transitions and associated criticality.

\end{abstract}

\maketitle

\section{Introduction}
The physics of nucleation, where a droplet having an order parameter value defining the global energy minima grows or shrinks within a metastable phase is one of the defining tenets within the study of phases and phase transitions~\cite{binder_rpp_1987,Chaikin_Lubensky_1995,livi_Politi_2017}. These ideas govern a wide range of phenomena spanning condensed matter~\cite{fokin_2006,kelton_book_2010,katsuno_pre_2011,vitelli_natcomm_2015}, nuclear physics~\cite{chung_jopGN_1993,strumia_nuphyb_1999,strumia_npb_1999,ali_prd_2022, MHindmarsh_1995, AresPRD22, Giombi_2024} and high-energy field theories~\cite{enghoff_grl_2011,eto_jhep_2022,ekstedt_jhep_2022,tranberg_jhep_2022,lofgren_prl_2023,hirvonen_2024,ashoke_sen_jhep_2015}. The conditions on growth and decay of any such nuclei, however, are governed by a competition between surface tension (interfacial line energy for a one-dimensional interface in a two-dimensional system) and bulk free-energy density~\cite{becker_doring_1935,gunton_book_1983,debenedetti_2020}. Microscopic models have been devised to make quantitative predictions and study physical processes such as crystal growth, adsorption, domain formation and role of thermal excitations, anisotropy in the domains and impurity effects~\cite{avron_jpamg_1982,akutsu_jpamg_1986,rikvold_pre_1994,acharyya_epjb_1998,akutsu_ptp_2001,akutsu_prb_2001,shneidman_prl_2006,ryu_pre_2010,katsuno_pre_2011}. With advances in computational methods and new experimental platforms, many of these ideas have been investigated ranging from material to colloidal systems~\cite{teeffelen_prl_2008,savage_prl_2009,sanz_acs_2013,herlach_jocp_2016,dudek_aicis_2020,suzuki_crengcom_2025}. While dominantly studied in classical systems, the nucleation physics in presence of quantum fluctuations and topological fermions have been little explored~\cite{fialko_prl_2012,yuval_prl_2022}. Insights from these might be crucial to understand domain formation and nucleation processes in interacting topological insulators where Landau-like order parameters interplay with band topology~\cite{rachel_rpp_2018, grover_nature_2022}.

\begin{figure}
\centering
\includegraphics[width=0.8\linewidth]{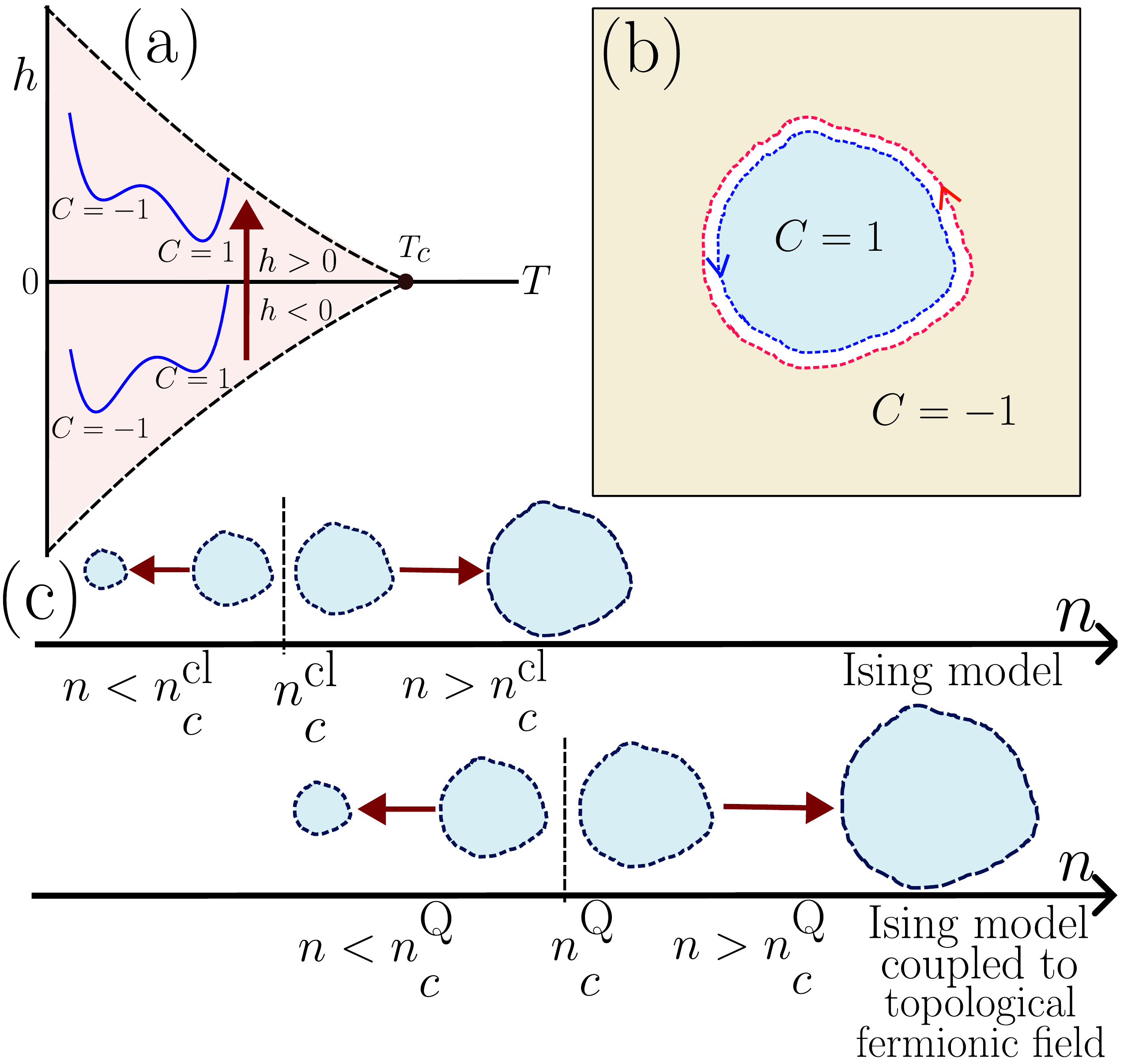}
\centering
\caption{{\textbf{Topological droplet:}} (a) Free-energy density (blue curves) for a two-dimensional classical Ising model coupled to topological fermions in magnetic field $h$ and temperature $T$ where $C$ is Chern number and arrow shows sudden quench. (b) $C=1$ droplet within a metastable region of $C=-1$. (c) A droplet of size $n$ either grows or shrinks depending on a critical size. This critical size for classical Ising system ($n_{c}^{\rm{cl}}$) increases to $n_{c}^{\rm {Q}}$ when coupled to topological fermions due to droplet edge-modes leading to additional interfacial line energy.}
\label{fig_schematic}
\end{figure}

In this work we pose and solve the question -- can a topological fermionic field, when coupled to a scalar field, modify the physics of nucleation? We answer this in affirmative, uncovering the role of topological edge-modes in modifying the interface energy and nucleation physics. Such edge-modes are a hallmark of topological phases of matter where they retain their robustness even when the bulk is gapped
\cite{hasan_rmp_2010,qi_rmp_2011,chiu_rmp_2016,bernevig_nature_2017}.  While our investigation here is theoretical, such systems of coupled fermionic degrees of freedom with Landau-like scalar order parameters have been of immense interest given their direct applicability to a range of material platforms including those of anomalous quantum Hall systems, spintronics and physics of correlated topological phenomena~\cite{simon_prl_2007,wang_apl_2014,sinova_revmodphys_2015,smejkal_prl_2017,smejkal_nature_2018,huber_natcomm_2020,bonbien_jpdap_2021,yang_prb_2023,zhu_apr_2023,ara_prl_2012,yang_prl_2014,krempa_arcmp_2014,amaricci_prb_2018}.

We limit our study to a scalar Ising field in two dimensions which undergoes the standard order-disorder transitions in the magnetic field ($h$) and temperature ($T$) plane (see Fig.~\ref{fig_schematic}(a)) where the second order thermal phase transition happens at $T=T_c$~\cite{onsager_physrev_1944}. We couple a two dimensional fermionic Chern insulator to the Ising field in such a way that the order parameter of scalar field acts like effective mass for the Dirac fermions. Thus the free-energy minima for $h>0$ and $h<0$ (for $T<T_c$) regime also correspond to topological insulating phases (for fermions) albeit with different Chern numbers ($C=\pm 1$). We perform a gedanken experiment where a quench is done from $h<0$ to $h>0$ regime such that the system is in a metastable phase of $C=-1$ with spins down polarized while the global free-energy minima is for $C=+1$ phase with spins up polarized. We now investigate the stability of a bubble of $C=+1$ region created within this metastable phase (see Fig.~\ref{fig_schematic}(b)). In absence of any fermionic field it is known that a bubble only beyond a critical size $n_{c}^{\rm{cl}}$ grows, while droplets of size $n<n_{c}^{\rm{cl}}$ shrinks with time. Here, we demonstrate that presence of a topological fermionic field when coupled to a scalar field leads to quantum corrections to the interface energy per unit length (also known as interfacial line energy) essentially due to topological edge-modes on droplet boundary (see Fig.~\ref{fig_schematic}(b)). This quantum correction to the interfacial line energy leads to an enhancement of the critical nuclei size to $n_{c}^{\rm{Q}}$ such that $n_{c}^{\rm{Q}}$ is always greater than $n_{c}^{\rm{cl}}$ (see Fig.~\ref{fig_schematic}(c)). In fact as we will show such enhancement can be tuned by changing the microscopic parameters of theory, easily achieving $n_{c}^{\rm{Q}}/n_{c}^{\rm{cl}} \sim 2$ within experimentally relevant scales. 

\section{Model}
We consider Bernevig-Hughes-Zhang (BHZ) model~\cite{bhz_2006} of spinless fermions coupled to two-dimensional ferromagnetic classical Ising model on a square lattice, such that the Hamiltonian is
\begin{multline}\label{eq_hbhz}
H=- \kappa \sum_{i} s_{i} {\bf{\Psi}}_{i}^{\dagger} \sigma_{z} ~{\bf{\Psi}}_{i} - \sum_{\langle ij \rangle} ({\bf{\Psi}}_{i}^{\dagger} {\large{\eta}}_{ij}~{\bf{\Psi}}_{j} + {\rm{h.c.}}) \\-J\sum_{\langle ij \rangle} s_{i} s_{j} -h \sum_{i} s_{i},
\end{multline}
where the spin variable for $i$-th unit cell can assume $s_{i}=\pm 1$ and $J>0$. Each unit cell contains $A$ and $B$ sites having staggered masses $\pm \kappa s_{i}$. The hopping strengths between nearest neighboring unit cells along $x$-axis and $y$-axis are ${\large{\eta}}_{ij}=\frac{1}{2} (\sigma_{z}+i \sigma_{x})$ and ${\large{\eta}}_{ij}=\frac{1}{2} (\sigma_{z}+i \sigma_{y})$ respectively where $\sigma_{x}$, $\sigma_{y}$, $\sigma_{z}$ are Pauli matrices and
${\bf{\Psi}}_{i}=\begin{pmatrix}
c_{iA} & c_{iB}
\end{pmatrix}^{\rm{T}}$
with $c_{iA}$ ($c_{iB}$) being annihilation operator of fermion at site $A$ ($B$) of $i$-th unit cell.
$\kappa$ is coupling parameter between the fermions and Ising spins.

\begin{figure}
\centering
\includegraphics[width=1.0\linewidth]{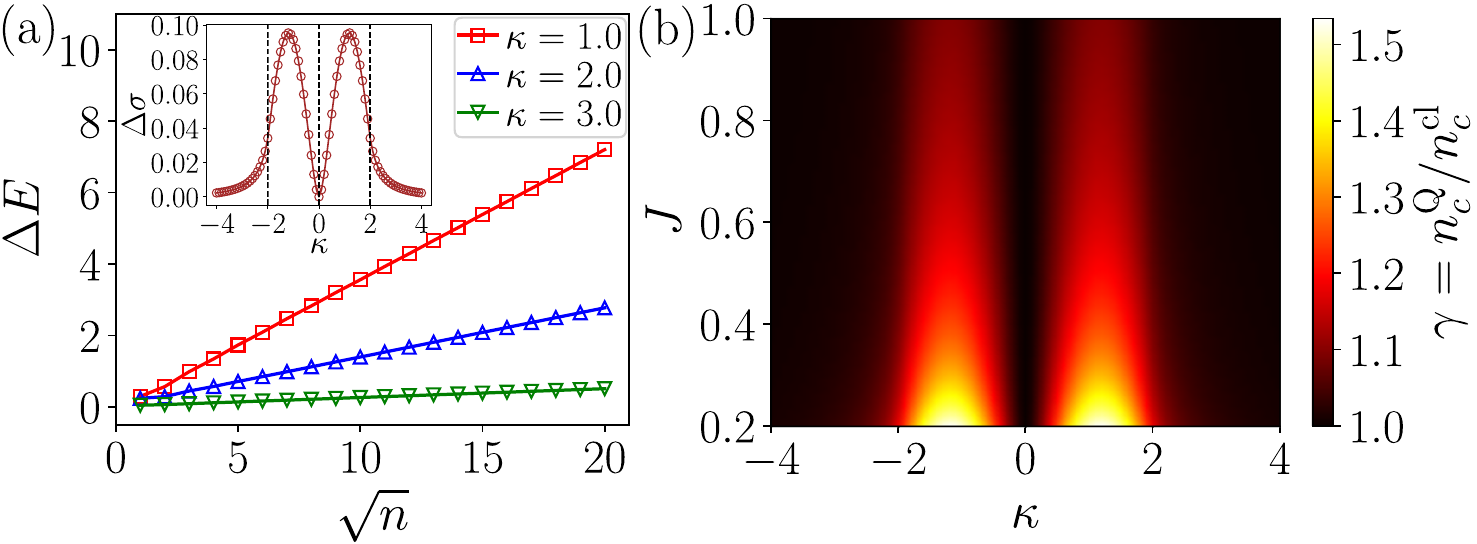}
\centering
\caption{{\textbf{Interface energy and critical cluster-size:}} (a) Change of ground-state energy $\Delta E$ for fermionic system with $\sqrt{n}$ where $n$ is cluster-size. (Inset) Correction to interfacial line energy $\Delta \sigma$ obtained from slope of linear fit (see Eq.~\eqref{eq_dE}) in (a) as a function of $\kappa$, where dotted lines denote quantum critical points $\kappa=-2,0,2$. (b) $\gamma=n_{c}^{\rm{Q}}/n_{c}^{\rm{cl}}$ (see Eq.~\eqref{eq_gamma}) with $\kappa$ and $J$ at $T=0$, where $n_{c}^{\rm{cl}}$ is critical cluster-size for classical situation ($\kappa=0$). Here, number of unit cells is $L^{2}=4096$. In topological situation ($0<|\kappa|<2$), $n_{c}^{\rm{Q}}>n_{c}^{\rm{cl}}$ and relative enhancement $\gamma$ increases with decreasing $J$.}
\label{fig_zeroT}
\end{figure}

We note that for a given $\{s_{i} \}$, the fermionic Hamiltonian in Eq.~\eqref{eq_hbhz} does not have time-reversal and sub-lattice symmetries. However, it preserves the following symmetries:
\begin{enumerate} 
\item Spin-fermion flip symmetry: This corresponds to the symmetry operation \text{$s_i \rightarrow -s_i,
{\bf{\Psi}}_i \rightarrow   (-1)^{i_x + i_y} \sigma _x {\bf{\Psi}}_i \otimes {\mathcal{K}}$} where ${\mathcal{K}}$ is the complex conjugation and $(i_{x}, i_{y})$ is the position of $i$-th unit cell.
\item Particle-hole (fermion) symmetry: Under this operation ${\bf{\Psi}}_i \rightarrow \sigma_x {\bf{\Psi^\dagger}}_i^T $, $s_i \rightarrow s_i$, $H \rightarrow H$, rendering the fermionic Hamiltonian into the symmetry class D.
\end{enumerate}

When $\kappa =1$, an up polarized state with magnetization $m = \frac{1}{L^2} \sum_i \langle s_i \rangle = +1$  corresponds to a fermionic Chern insulator with $C=+1$ and a down polarized state ($m= -1$) leads to $C=-1$. Moreover at $T=0$,  $\kappa$ itself can be tuned such that when $0<|\kappa|<2$ we have a topological phase with $C=\pm 1$ and a trivial phase with $C=0$ when $|\kappa|>2$~\cite{bhz_2006,Bernevig_book_2013,asboth_2016}. We will consider periodic boundary conditions throughout our work.

Let us now consider the system being subjected to a sudden quench of external field from $h<0$ to $h>0$ (keeping $|h|$ fixed) at a fixed temperature $T < T_{c}$. The system thus forms a metastable state of $m=-1$, whose stability to small droplets of $m=+1$ (the global minima) can be analyzed, as we discuss below.

\section{Topological interface energy}
When $\kappa=0$, the fermions and spins are decoupled and the physics of nucleation under the quench protocol is determined by just the classical Ising fields. Thus, change of free-energy in the formation of a circular cluster of size $n$ is 
\begin{equation}
\Delta F_{\rm{classical}} = -2|h| n + 2 {\sqrt{\pi n}} \sigma_{\rm{cl}}, 
\end{equation}
where $\sigma_{\rm{cl}}>0$ is the interfacial line energy which depends on $J$ and temperature $T$~\cite{avron_jpamg_1982,akutsu_jpamg_1986,Chaikin_Lubensky_1995,kevin_pre_2005}.  When $\kappa=0$, change of free energy $\Delta F_{\rm{classical}}$ is maximum for critical cluster-size 
\begin{equation}
n_{c}^{\rm{cl}}=\frac{\pi \sigma_{\rm{cl}}^{2}}{4 |h|^{2}}.
\end{equation}
Thus, to reduce the free energy, the cluster grows with time when $n>n_{c}^{\rm{cl}}$ and it shrinks when $n<n_{c}^{\rm{cl}}$, determined by a competition of the interfacial line energy with the bulk energy-density and a system with a higher $\sigma_{\rm{cl}}$ would pertain to a larger $n_{c}^{\rm{cl}}$.

When $\kappa \neq 0$, the Ising field and the fermionic system are coupled thus changing the total free-energy of the system. Under the same quench protocol, total change of free-energy $\Delta F$ is now
\begin{equation}\label{freeEn}
\Delta F=\Delta F_{\rm{classical}} +\Delta E,
\end{equation}
where $\Delta E$ is the quantum contribution from fermionic field which can again have both the interface and bulk contributions. However, due to the spin-fermion flip symmetry, the energy eigenvalues $E_{j}$ of bulk spectrum for a given $\{s_{i} \}$ are independent of the sign of $\kappa$, i.e.~$E_{j} (\kappa) = E_{j} (-\kappa)$. This also reflects into a spin-flip symmetry as $E_j (\kappa, \{s_i\}) = E_j (\kappa, \{-s_i\})$.  This guarantees that under $s_{i} \to -s_{i}$, the change in the bulk energy density of fermionic system $\Delta \epsilon =0$. In other words, the bulk contribution from the fermionic field is zero since in the nucleating droplet there is a flip of the effective mass term $M_{\text{eff}}\equiv \kappa m$ under which the ground state energy of fermionic system at half-filling does not change. The interfacial contribution to $\Delta E$ can however be non-zero depending on the character of the fermionic phases as we discuss below. 

To estimate $\Delta E$ we compare the ground state (i.e. at half-filling) energies ($h=0, T=0$) between configurations (i) where a square region of size $L^2$ has all $s_i=-1$ and another (ii) where an internal square-shaped region of size $l^2$ (where there are $l=\sqrt{n}$ unit cells along both $x$ and $y$-axis) has $s_{i}=+1$ while the rest of $s_{i}=-1$. Variation of $\Delta E$ shows a characteristic $\sqrt{n}$ dependence for a finite $\kappa$ (see Fig.~\ref{fig_zeroT}(a)), thus leading to a quantum contribution to the interfacial line energy ($\Delta \sigma$)~\cite{tanaka_prl_2022,tanaka_prb_2023} which can be estimated from the slope as 
\begin{equation}\label{eq_dE}
\Delta E = 4 (\Delta \sigma) \sqrt{n},
\end{equation}
where $4 \sqrt{n}$ is the perimeter of the internal region. Furthermore $\Delta \sigma$ has an interesting dependence on $\kappa$ which we discuss next.

As shown in inset of Fig.~\ref{fig_zeroT}(a), $\Delta \sigma \rightarrow 0$ when $\kappa \rightarrow 0$ reflecting the decoupled limit. However $\Delta \sigma$ is significantly larger in topological situation ($0<|\kappa|<2$) compared to the trivial situation ($|\kappa|>2$). In the topological situation the droplet necessarily hosts robust edge-modes at the boundary due to a change in the Chern number between the cluster and its  background. This imposes a reorganization of the energy-spectrum, causing fermions to move up from the bulk energy levels to mid-gap edge modes. Given a bulk gap of $\sim \Delta_{\rm{bg}}$, the correction to the interfacial line energy in the topological situation is primarily determined by the edge-mode energy per unit length of the boundary of the droplet (see Appendix~\ref{app_source}):
\begin{equation}
    \Delta {\mathcal{E}}_{\rm{edge}} \sim  \frac{\Delta_{\rm{bg}}^2}{16 \pi \hbar v_F}
\end{equation}
where $v_F$ is the Fermi velocity of the edge modes. This should be contrasted with the trivial situation where no such edge-mode appears resulting in $\Delta \sigma \rightarrow 0$. Thus, the role of edge-modes as the primary source of the interfacial line energy in the topological phase is established. In general, a nucleating domain of one fermionic phase of matter into another, even both being trivially insulating, has a non-zero interface energy contribution due to perturbative reorganization of all bulk states resulting from domain boundary. Such contributions are, however, non-universal and much smaller than those from topological edge modes. A demarcation of such contributions is discussed in Appendix~\ref{app_source}.

\begin{figure}
    \centering
\includegraphics[width=1.0\linewidth]{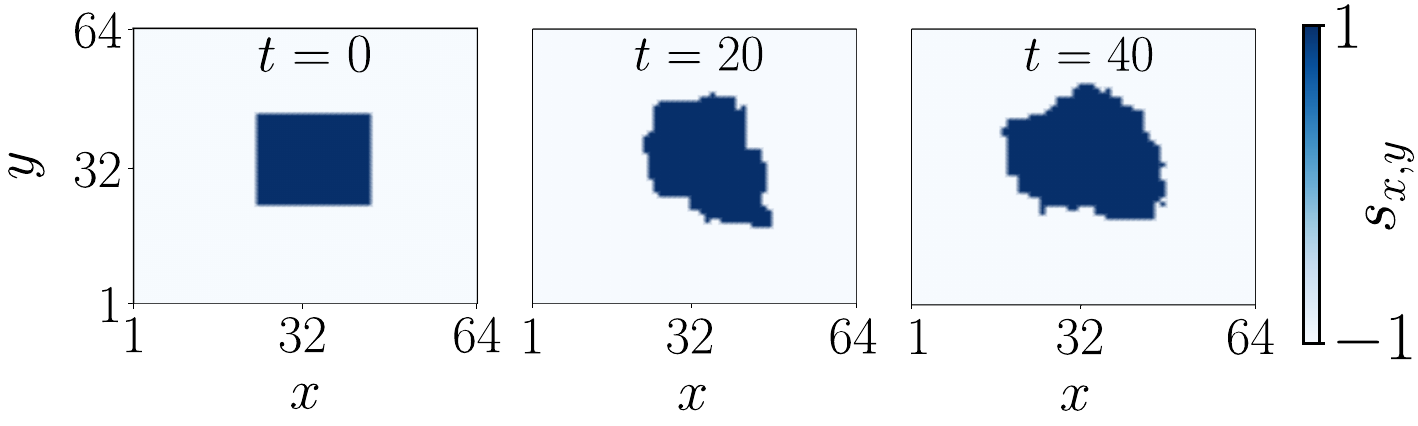}
\caption{
{\textbf{Local magnetization in coupled spin-fermion Monte Carlo simulation:}} Local magnetization $ s_{x,y}$ for unit cell with position $(x,y)$ at times $t=0,20,40$ (in units of Monte Carlo steps per site) in a typical realization of the coupled spin-fermion exact Monte Carlo simulation (see Eq.~\eqref{eq_diff_qme} and Eq.~\eqref{eq_qm_mc}) for $\kappa=1.2$. Here, the initial cluster-size with up-spins is $n=441$. The parameters chosen are: $J=0.2$, $T=1/\beta=0.33$ (where $T_{c} \approx 0.45$), $h=0.02$ and number of unit cells in square lattice $L^{2}=4096$. We observe the growth of the cluster.}
\label{fig_exact_mc}
\end{figure}

\begin{figure*}
    \centering
\includegraphics[width=1.0\linewidth]{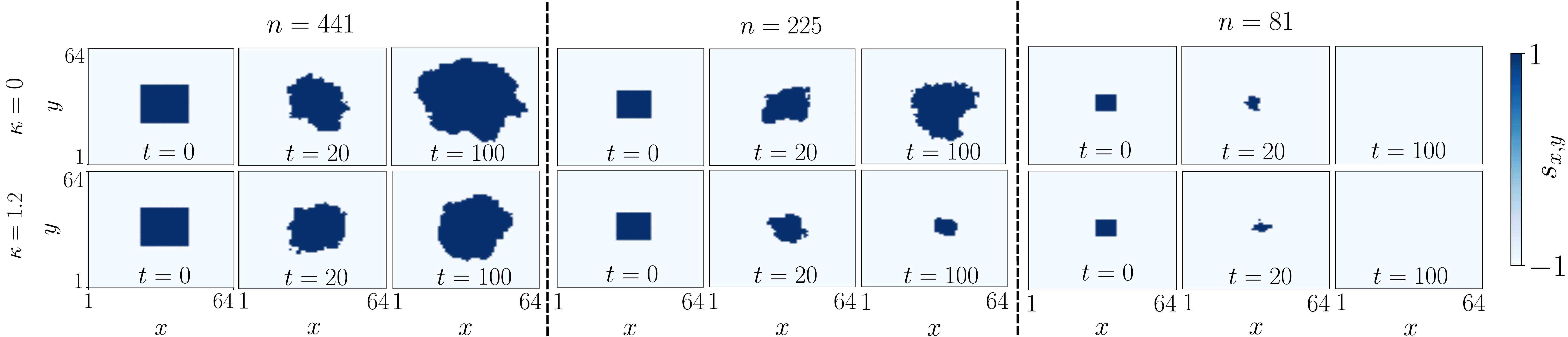}
\caption{{\textbf{Local magnetization in a typical realization:}} Local magnetization $ s_{x,y}$ for unit cell with position $(x,y)$ at times $t=0, 20, 100$ (in units of Monte Carlo steps per site) in a typical realization for $\kappa=0$ and $\kappa=1.2$ where initial cluster-sizes with up-spins are $n=441$, $n=225$ and $n=81$. The parameters chosen are: $J=0.2$, $T=1/\beta=0.33$ (where $T_{c} \approx 0.45$), $h=0.02$. Number of unit cells in square lattice is $L^{2}=4096$. The cluster with initial size $n=441$ grows in both the decoupled ($\kappa=0$) and topological ($\kappa=1.2$) situations, while the cluster with $n=81$ shrinks in both the situations. When $n=225$, we observe the growth of the cluster for $\kappa=0$ and the shrinkage of the cluster for $\kappa=1.2$.}
\label{snapshot_supplemental_fig}
\end{figure*}

When $\kappa \neq 0$, total change of free energy $\Delta F$ is maximum for the critical circular cluster-size
\begin{equation}
n_{c}^{\rm{Q}}=\frac{\pi \left(\sigma_{\rm{cl}} + \Delta \sigma \right)^{2}}{4 |h|^{2}}.
\end{equation}
Thus the enhancement of critical nuclei of classical Ising field, due to its coupling to fermionic field ($\kappa \neq 0$) can be quantified by defining 
\begin{equation}\label{eq_gamma}
 \gamma =\frac{n_{c}^{\rm{Q}}}{n_{c}^{\rm{cl}}}=\left(1+\frac{\Delta \sigma}{ \sigma_{\rm{cl}}}\right)^{2}.
\end{equation}
Interestingly, $\gamma$ behaves non-monotonically with $\kappa$, mirroring the behavior of $\Delta \sigma$. The interplay of Ising spin-exchange scale $J$ and edge-mode contribution to $\Delta \sigma$ allows for further tunability of $\gamma$. At $T=0$, the interfacial line energy in classical Ising model with square-shaped droplet boundary is $\sigma_{\rm{cl}} \sim 2J$, while the exact value of $\sigma_{\rm{cl}}$ depends on the anisotropy of the droplet and the direction of its growth~\cite{avron_jpamg_1982,akutsu_jpamg_1986}. 
We observe an increase in $\gamma$ with decreasing $J$ in topological situation ($0<|\kappa|<2$). For $J \sim 0.2$, $\kappa \sim 1$, one can tune $\gamma \sim 1.5$ which is upto $50\%$ rise in the size of critical nuclei; while in trivial situation ($|\kappa|>2$), $\gamma \sim 1$, implying no significant change in critical cluster-size (see  Fig.~\ref{fig_zeroT}(b)).

While our discussion of topological interface energy has been done in context of model introduced in Eq.~\eqref{eq_hbhz}, the phenomenology holds true in general. Due to the spin-fermion flip symmetry for the model in Eq.~\eqref{eq_hbhz}, the fermionic volume contribution to free energy $\Delta \epsilon=0$ where $\Delta E \propto \Delta \epsilon n$ (see near Eq.~\eqref{freeEn}). The quench physics then is solely determined by the interfacial contribution. However, in absence of this symmetry, when $\Delta \epsilon \neq 0$, the topological interface energy contribution still dominantly governs the nucleation physics in topological regime albeit with renormalized $n^{\rm{Q}}_{c}$ and $\gamma$ (see Appendix~\ref{app_symmetry}). To further study if this physics remains stable under thermal fluctuations, we now analyze the system at finite temperatures.

\section{Growth of clusters at finite temperature}
At finite $T$, thermal fluctuation reduces the stiffness of domain wall thereby reducing $\sigma_{\rm{cl}}$ from its $T=0$ value ($\sim 2J$)~\cite{shneidman_jcp_1999,kevin_pre_2005}. It is useful to note we work with $T< T_c$ where classical transition temperature $T_{c} \approx 2.269 J$~\cite{onsager_physrev_1944} (where Boltzmann constant is set to unity). Moreover, we consider temperatures much smaller than fermionic energy-scales, thus the effect of thermal fluctuation on $\Delta \sigma$ can be ignored. Thus with increasing temperature, since effective $\sigma_{\rm{cl}}$ reduces, we expect that $\gamma$ increases (see Appendix~\ref{app_thermal}). 

To analyze the critical cluster size and growth of the clusters at finite $T$, we resort to Monte Carlo (MC) simulation and employ the Metropolis algorithm~\cite{metropolis_jcp_1953}. When the spins and fermions are decoupled ($\kappa=0$), a random spin $s_{i}$ is chosen and the energy $\Delta E_{i}$ required to flip the spin (i.e., $s_{i} \to -s_{i}$) is calculated as:
\begin{equation}
\Delta E_{i}=2Js_{i} \sum_{j \in {\mathcal{N}}_{i}} s_{j}+2h s_{i},
\end{equation}
where $j$ runs over all the nearest neighbors (${\mathcal{N}}_{i}$) of $i$-th spin. Now, if $\Delta E_{i}<0$ or if $p_{i} \leq \exp(-\beta \Delta E_{i})$ for a randomly chosen $p_{i} \in [0,1]$ where $\beta=1/T$, the spin flip is accepted and the $i$-th spin becomes $-s_{i}$. If none of these conditions is satisfied, the spin-flip is rejected and $i$-th spin remains $s_{i}$.

When the classical Ising field is coupled to the fermionic field ($\kappa \neq 0$), both the spins and fermions contribute to total energy. Here we solve the fermionic problem to compute $\Delta E_{i}$ at each step of Monte Carlo simulation. In this method, we propose a spin flip $s_{i} \to -s_{i}$ at a randomly chosen $i$-th unit cell at each Monte Carlo step. We then explicitly calculate the difference $\Delta E_{i}^{\rm{Q}}$ in the ground state energies between the configurations after and before the proposed $i$-th spin-flip for the fermionic problem as:
\begin{equation}\label{eq_diff_qme}
\Delta E_{i}^{\rm{Q}}=E^{\rm{Q}}\left(s_{1},..., -s_{i},...,s_{L^{2}} \right)-E^{\rm{Q}} \left(s_{1},...,s_{i},...,s_{L^{2}} \right),
\end{equation}
where $E^{\rm{Q}}(s_{1},...,s_{i},...,s_{L^{2}})$ denotes the ground state energy for the fermionic system with spin-configuration $s_{1},...,s_{i},...,s_{L^{2}}$. Thus, the total energy required for the $i$-th spin-flip is given by:
\begin{equation}\label{eq_qm_mc}
\Delta E_{i}=2Js_{i} \sum_{j \in {\mathcal{N}}_{i}} s_{j}+2h s_{i} + \Delta E_{i}^{\rm{Q}}.
\end{equation}
Eq.~\eqref{eq_qm_mc} can be used to check the acceptance or rejection of the spin-flip at each Monte Carlo step when $\kappa \neq 0$. Under this protocol, the spin fluctuations and their effect on fermions are exactly accounted for.

\begin{figure*}
    \centering
\includegraphics[width=1.0\linewidth]{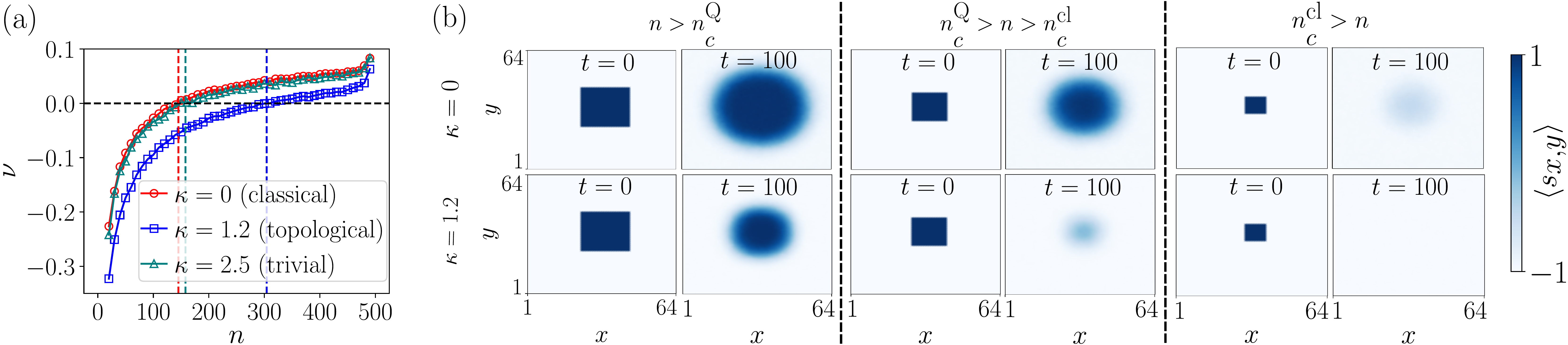}
\caption{{\textbf{Critical cluster-size and droplet growth:}} (a) $\nu$ as a function of cluster-size $n$ for 
classical ($\kappa=0$), 
topological ($\kappa=1.2$) and  trivial ($\kappa=2.5$) situations (see Eq.~\eqref{eq_nu}). $\nu =0$ denotes the critical sizes which are $n_{c}^{\rm{cl}}=145$, $n_{c}^{\rm{Q}}=304$, $n_{c}^{\rm{Q}}=158$ respectively. While in topological situation $n_{c}^{\rm{Q}}>n_{c}^{\rm{cl}}$, in trivial situation $n_{c}^{\rm{Q}} \approx n_{c}^{\rm{cl}}$.
(b) Initial and final average local magnetization $\langle s_{x,y} \rangle$ when clusters of sizes, $n=441$ ($n>n_{c}^{\rm{Q}}$), $n=225$ ($n_{c}^{\rm{Q}}>n>n_{c}^{\rm{cl}}$), $n=81$ ($n_{c}^{\rm{cl}}>n$) are evolved. The final time $t=100$ is in units of MC steps per site. Upper (lower) panel shows $\kappa=0$ ($\kappa=1.2$) and $J=0.2$, $T=1/\beta=0.33 < T_{c} \approx 0.45$, $h=0.02$, $L^{2}=4096$. Averaging is performed over $10^3$ simulations.  For classical ($\kappa=0$) situation, the cluster grows when $n>n_{c}^{\rm{cl}}$ but in topological  situation ($\kappa=1.2$), growth occurs only when $n>n_{c}^{\rm{Q}}$.}
\label{snapshot_classical_quantum}
\end{figure*}

We perform an exact MC simulation for $\kappa \neq 0$, starting from an initial configuration with a cluster of up-spins and size $n=l^{2}$ within the region of all down-spins. For a typical realization, the local magnetization ($s_{x,y}$) for any unit cell at position ${\textbf r} =(x,y)$ as a function of time $t$ (in units of Monte Carlo steps per site or after $tL^{2}$ Monte Carlo steps where $L^{2}$ is the number of spins in square lattice) illustrates the growth of the cluster (see Fig.~\ref{fig_exact_mc}). The exact MC simulation, however, is numerically expensive, since at every MC step, a matrix diagonalization for the fermionic system needs to be performed. This step can be simplified by devising an effective Monte Carlo method where the electronic effects can be effectively included as internal energy corrections to the spin configurations within Metropolis algorithm (see Appendix~\ref{app_monte_carlo}). This simplification follows from the fact that the electronic time-scales are much faster than the spin time-scales. As shown in Fig.~\ref{snapshot_supplemental_fig}, the effective MC simulation can capture the growth (shrinkage) of the clusters with various initial sizes $n$ for both the decoupled ($\kappa=0$) and topological ($\kappa=1.2$) situations. We also observe that the boundaries of the clusters follow very rugged trajectories during their growth or shrinkage.

In order to understand the behavior of the growth (shrinkage) of the clusters for the decoupled and topological situations, we proceed to the numerical evaluation of the critical cluster sizes $n_{c}^{\rm{cl}}$ and $n_{c}^{\rm{Q}}$ using the effective MC simulation. Here, a cluster of up-spins and finite size $n$ is placed within the metastable background of down-spins and is subjected to a non-equilibrium situation (due to a sudden quench of $h$). Considering various realizations of the simulation, we then investigate the quantity
\begin{equation}\label{eq_nu}
\nu(n)=\frac{P_{+}(n)-P_{-}(n)}{{P_{+}(n)+P_{-}(n)}}
\end{equation}
where $P_{+}(n)$ and $P_{-}(n)$ count the number of MC steps in which size $n$ of a cluster increases to $(n+1)$ and decreases to $(n-1)$ respectively~\cite{katsuno_pre_2011}. When the cluster-size $n$ is smaller than the critical cluster-size, the tendency of shrinkage of the cluster is greater than that of growth, implying $P_{+}<P_{-}$ and $\nu<0$, while when $n$ is greater than the critical cluster-size, the tendency of growth of the cluster is greater than that of shrinkage, implying $P_{+}>P_{-}$ and $\nu>0$. Therefore, when $n$ is equal to the critical cluster-size, $\nu=0$. Critical cluster-sizes $n_{c}^{\rm{cl}}$ for classical situation ($\kappa=0$) and $n_{c}^{\rm{Q}}$ for both topological ($\kappa=1.2$) and trivial ($\kappa=2.5$) 
 situations are thus obtained from when $\nu$ crosses $0$, where we find $n_{c}^{\rm{Q}}>n_{c}^{\rm{cl}}$ in topological situation and $n_{c}^{\rm{Q}} \approx n_{c}^{\rm{cl}}$ in trivial situation (see Fig.~\ref{snapshot_classical_quantum}(a)) thus showing the characteristic rise of the interfacial line energy only in the topological case.

We then move on to explore the universal feature of growth (shrinkage) of the clusters through the dynamics of the local magnetization averaged over various realizations. The behavior of the average local magnetization $\langle s_{x,y} \rangle$ for both $\kappa=0$ and $\kappa=1.2$ is shown in Fig.~\ref{snapshot_classical_quantum}(b). Unlike the rugged boundaries of the clusters in a typical realization, the averaging of the local magnetization over various realizations renders the clusters to appear circularly symmetric. We find that when initial size of cluster (having up-spins) $n>n_{c}^{\rm{Q}}$, the cluster grows with time $t$ and 
when $n_{c}^{\rm{cl}}>n$, the cluster shrinks. This behavior remains unchanged in both the classical ($\kappa=0$) and the topological ($\kappa = 1.2$) situations. However, when 
$n_{c}^{\rm{Q}}>n>n_{c}^{\rm{cl}}$, although the cluster grows for $\kappa=0$, it shrinks for $\kappa=1.2$ (see Fig.~\ref{snapshot_classical_quantum}(b)). These observations establish the role of the topological interface energy contribution in altering growth (shrinkage) of clusters.

While we have established the role of topological interface energy contribution in affecting the nucleation physics, modifying the quench protocols can allow further tunability in such growth processes (see Appendix~\ref{slow}).

\section{Edge-state evolution}
\begin{figure}
    \centering
\includegraphics[width=1.0\columnwidth]{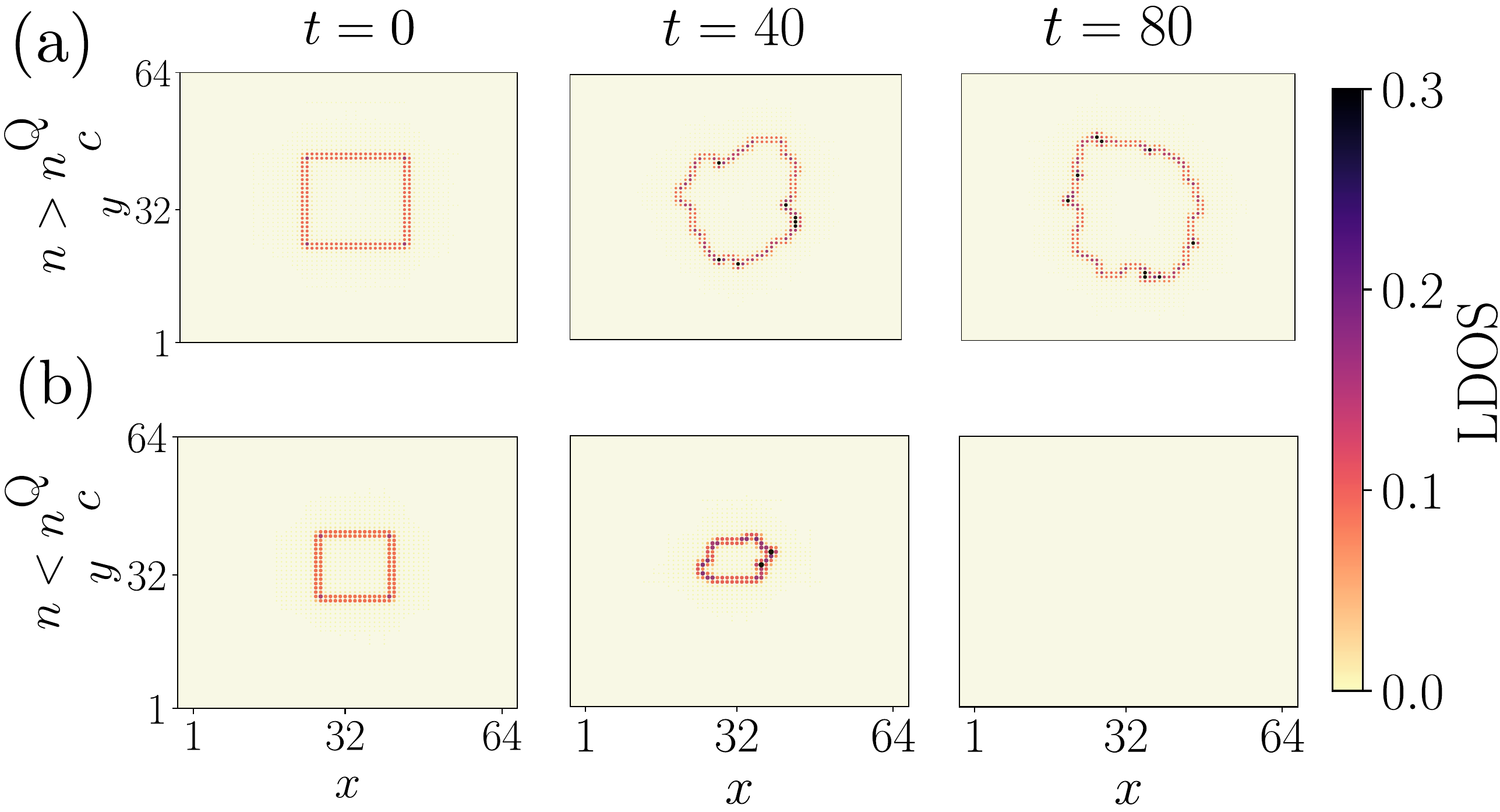}
    \caption{
    {\textbf{Edge-state evolution:}} Local density of states (LDOS) of edge-modes for initial cluster-sizes (a) $n=441$ ($n>n_{c}^{\rm{Q}}$), (b) $n=225$ ($n<n_{c}^{\rm{Q}}$) at times $t=0,40,80$ (in the units of Monte Carlo steps per site), where $n_{c}^{\rm{Q}}=304$. The parameters chosen here are: $J=0.2$, $T=1/\beta=0.33$ (where $T_{c} \approx 0.45$), $h=0.02$, $\kappa=1.2$. The number of unit cells in square lattice is $L^{2}=4096$. We observe that LDOS of edge-modes captures the signatures of growth (when $n>n_{c}^{\rm{Q}}$) and shrinkage (when $n<n_{c}^{\rm{Q}}$) of clusters for $0<|\kappa|<2$.}
    \label{edge_prob}
\end{figure}
In order to observe the edge-states in a nucleating droplet, we take a typical spin configuration and solve the fermionic problem at every step in the MC evolution to obtain the local density of states (LDOS) defined as
\begin{equation}
{\rm{LDOS}}=\sum_{j} \left( |\langle i,A | \psi_{j} \rangle|^{2} + |\langle i,B | \psi_{j} \rangle|^{2} \right),
\end{equation}
at $i$-th unit cell with coordinate $(x,y)$ (where $j$ runs over all eigenstates corresponding to single-particle energy-eigenvalues $|E_{j}| < 0.05 W$ near the Fermi energy where $W$ is the bandwidth of single-particle spectrum) at various times $t$. Unsurprisingly the edge-state evolution in topological situation ($0<|\kappa|<2$) follows the signatures of growth and shrinkage of clusters. We find that edge-modes are indeed localized at the boundary between two different topological regions in any typical realization (see Fig.~\ref{edge_prob}). When $n>n_{c}^{\rm{Q}}$, the boundary of cluster grows with time $t$, while it shrinks and eventually vanishes when $n<n_{c}^{\rm{Q}}$ (see Fig.~\ref{edge_prob}(a,b)). Taken together, all these results show a characteristic interfacial line energy which can be alluded to a topological phase.

\section{Outlook}
In this work, we investigate nucleation of a scalar field when coupled to a topological fermionic field. Taking a concrete example of a two-dimensional Ising model coupled to a Chern insulator, we find that fermionic quantum contribution leads to an additional correction to the interfacial line energy in a nucleating cluster. Interestingly, the coupling parameter of the two fields can serve as a new tunable parameter for controlling the critical cluster-size in nucleation, thus affecting its growth and shrinkage. Ranging from implications on fundamental aspects of coupled topological field theories, just nucleation physics may also play a role in formation of topological domains such as Chern mosaic structures seen in strongly correlated topological systems~\cite{grover_nature_2022}. Our work opens up a range of questions regarding physics of nucleation when coupled to higher dimensional topological systems where one can systematically pose the role of surface tension in modifying critical properties of the classical fields. The imminent breakdown of adiabaticity would result in interplay of quantum and thermal fluctuations leading to intriguing critical theories with characteristic exponents. Furthermore, the role of Goldstone fluctuations when the 
classical field has continuous symmetry can also be interesting. The study of spinodal decomposition and coarsening dynamics~\cite{cahn_hilliard_1958,bray_aip_1994} could be important future directions.\\

\begin{acknowledgements}
We acknowledge fruitful discussions with Amit Agarwal, Vijay B.~Shenoy, Diptarka Das, Sabyasachi Chakraborty, Soumya Sur, Subrata Pachhal, Rahul Singh. S.M. acknowledges support from PMRF Fellowship, India. AA acknowledges support from IITK Initiation Grant (IITK/PHY/2022010). Numerical calculations were performed on the workstations {\it Wigner} and {\it Syahi} at IITK. 
\end{acknowledgements}

\appendix

\section{Sources of Interface energy}\label{app_source}

We demarcate both the bulk and edge contributions to $\Delta \sigma$ for the model discussed in the main text.

\subsection{Calculation of edge mode energy}\label{app_edge_energy}
To compute edge-mode energy per unit length for BHZ model, we consider the Hamiltonian
\begin{equation}\label{eq_bhz_real}
H=- M \sum_{i} {\bf{\Psi}}_{i}^{\dagger} \sigma_{z} ~{\bf{\Psi}}_{i} - \sum_{\langle ij \rangle} ({\bf{\Psi}}_{i}^{\dagger} {\large{\eta}}_{ij}~{\bf{\Psi}}_{j} + {\rm{h.c.}}).
\end{equation}
For periodic boundary conditions in both $x$ and $y$ directions, we resort to momentum space where Hamiltonian for $k_{x}, k_{y} \in (-\pi, \pi]$ is
\begin{equation}
H_{k_{x},k_{y}}=\sin(k_{x})\sigma_{x} + \sin(k_{y})\sigma_{y} + (-M - \cos(k_{x}) - \cos(k_{y})) \sigma_{z}.
\end{equation}
Thus, the dispersion relation is given by
\begin{widetext}
\begin{equation}
{{E(k_{x}, k_{y}) =\pm \sqrt{\sin^{2}(k_{x}) + \sin^{2}(k_{y}) + (M + \cos(k_{x}) + \cos(k_{y}))^{2}}.}}    
\end{equation}
\end{widetext}

\begin{figure*}
\centering
\includegraphics[width=1.0\linewidth]{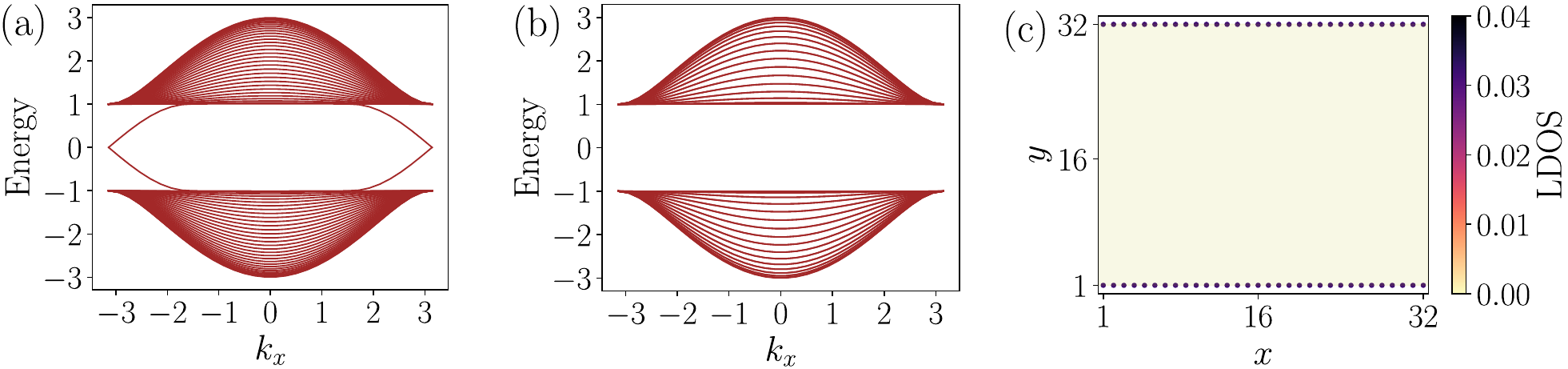}
\centering
\caption{{\textbf{Spectrum and edge modes in BHZ model:}} (a) Single-particle energy spectrum as a function of $k_{x}$ for BHZ model with $M=1.0$ for periodic boundary in $x$ direction and open boundary in $y$ direction, (b) single-particle energy spectrum as a function of $k_{x}$ for BHZ model with $M=1.0$ for periodic boundary in both $x$ and $y$ directions, (c) Local density of states (LDOS) of edge-modes at unit cell with coordinate $(x,y)$ for BHZ model with periodic boundary in $x$ direction and open boundary in $y$ direction for $L=32$, $M=1.0$.}
\label{fig:analytical_surface_tension}
\end{figure*}

Now, in a ribbon geometry with periodic (open) boundary condition in $x$ ($y$) direction, $k_{x}$ remains a good quantum number and hosts $x$ dispersing edge states on the top and bottom boundaries in the topological phase ($0<|M|<2$). The dispersion in the two situations (open and periodic boundary conditions) and the local density of states (LDOS) of the edge-states are shown in Fig.~\ref{fig:analytical_surface_tension}(a-c).

The difference in the ground-state energy for a half-filled system between the two situations (a) and (b) determines the mean edge-mode energy when the system is deep in the topological phase ($0 \ll |M| \ll 2$). For instance, at $M=1$ an edge state wavefunction that decays exponentially in the $y$ direction has an effective energy $E_{\rm{edge}}(k_{x}) \sim -\sin(k_x)$ near $k_x=\pi$ (see Fig.~\ref{fig:analytical_surface_tension}(a)). However, such an edge state can only be defined appropriately until it mixes with the bulk modes. The valence bulk band maxima $E_{\rm{valence}}(k_{x})$ is determined by $k_{y}=\pi$ mode when $0<M<2$ (similarly by $k_{y}=0$ mode when $-2<M<0$). Thus the range of $k_x$ until which the edge states survive is given by $k^{c}_{x}=\cos^{-1}(-M+1) = \pm \frac{\pi}{2}$ for $M=1$.
 The edge state energy is thus given by the difference 
\begin{equation}
\Delta E_{\rm{edge}}^{\rm{analytical}} 
= \frac{L}{\pi} \int_{ k^{c}_{x}}^{\pi} dk_{x} \left( E_{\rm{edge}}(k_{x}) - E_{\rm{valence}}(k_{x})  \right).
\end{equation}

For a $L \times L$ square lattice with periodic boundary condition in $x$ direction and open boundary condition in $y$ direction, the edge length $=2L$ and thus the edge-mode energy per unit length is 
\begin{align}\label{eq_edge_energy_density}
\Delta {\mathcal{E}}_{\rm{edge}} &= \frac{\Delta E_{\rm{edge}}^{\rm{analytical}}}{2L} \nonumber \\
&=\frac{1}{2\pi} \int_{k^{c}_{x}}^{\pi} dk_{x} \left( E_{\rm{edge}}(k_{x}) - E_{\rm{valence}}(k_{x})  \right),
\end{align}
which can be evaluated exactly and is $=\frac{(\pi-2)}{4 \pi}$ for $M=1$. For a general linearly dispersing edge state where the dispersion is given by
\begin{equation}
E_{\text{edge}}(k_{x}) \sim \hbar v_F k_x.
\end{equation}
For the bulk band-gap (energy difference between the valence band maximum to conduction band minimum) $\Delta_{\rm{bg}}$, $k_x^{c}$ can be estimated as $\sim \frac{\Delta_{\rm{bg}}}{2 \hbar v_F}$, thus 
\begin{equation}\label{eq_gap_est}
\Delta {\mathcal{E}}_{\rm{edge}} \sim  \frac{1}{2 \pi}\int_{0}^{\frac{\Delta_{\rm{bg}}}{2 \hbar v_F}} dk_{x} \hbar v_F k_{x}  \sim \frac{\Delta_{\rm{bg}}^2}{16 \pi \hbar v_F}.
\end{equation}
In the present Hamiltonian, since $v_F \sim 1$, and thus the interfacial line energy arising from the edge is expected to vary as the square of the bulk gap.

\subsection{Contribution of bulk and edge states to interface energy}

\begin{figure}
\centering
\includegraphics[width=1.0\columnwidth]{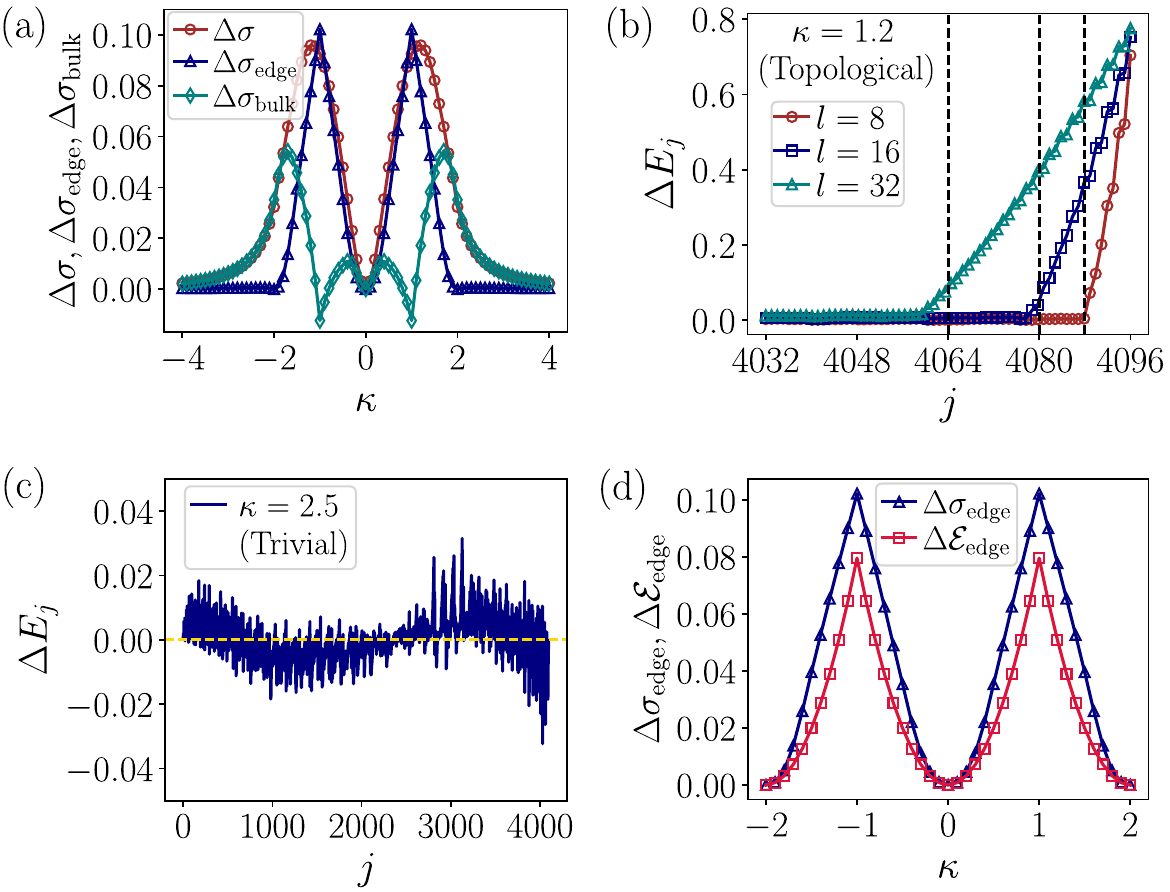}
\centering
\caption{
{\textbf{Contribution of edge-modes and bulk states to interfacial line energy:}} (a) Correction to the interfacial line energy $\Delta \sigma$, the contributions of bulk states $\Delta \sigma_{\rm{bulk}}$ and edge-modes $\Delta \sigma_{\rm{edge}}$ as a function of $\kappa$ where $\Delta \sigma_{\rm{bulk}}$ and $\Delta \sigma_{\rm{edge}}$ are obtained using Eq.~\eqref{eq_bulk_sigma} and Eq.~\eqref{eq_edge_sigma} respectively. (b) Difference of energy $\Delta E_{j}$ of $j$-th single-particle level (see Eq.~\eqref{eq_before_after}) below the Fermi energy in topological situation ($\kappa=1.2$) when the size of the droplet is $n=l^{2}$. (c) $\Delta E_{j}$ as a function of $j$ in trivial situation ($\kappa=2.5$) when $l=16$. (d) $\Delta \sigma_{\rm{edge}}$ and analytical estimate of $\Delta {\mathcal{E}}_{\rm{edge}}$ (see Eq.~\eqref{eq_gap_est}) as a function of $\kappa \equiv M$. In all plots, number of unit cells in square lattice is $L^{2}=4096$.
}
\label{fig:edge_bulk_surface_tension}
\end{figure}

 In order to isolate the contributions to the interfacial line energy due to the bulk and edge states separately, we devise the following numerical scheme. We compare the single particle energy eigenvalues between the two cases (i) a uniform square region of size $L^2$ with all $s_i=-1$ (i.e.~before quench), (ii) the case where within the region of size $L^{2}$, a square-shaped droplet of size $n=l^{2}$ with $s_{i}=1$ is now formed (i.e.~after quench). Each of these systems has $2 L^2$ single particle energy eigenvalues which can be enumerated and indexed in ascending order from $j=1, \ldots, 2L^2$. The difference in the single particle eigenvalues between the two cases is
\begin{equation}\label{eq_before_after}
    \Delta E_j = E_j (\text{after quench}) - E_j (\text{before quench})
\end{equation}
and their additive sum until the Fermi energy ($j=L^{2}$) is the essential contribution to $\Delta \sigma$ at temperature $T=0$ which was evaluated in the main text using 
\begin{equation}
    \Delta E = \sum_{j=1}^{L^2} \Delta E_j = 4 \Delta \sigma \sqrt{n}.
\end{equation}

Given we know that in the topological phase, $l$ eigenstates below the Fermi energy form the edge state manifold, the contribution to $\Delta \sigma$ from the bulk and edge can be separately estimated as
\begin{equation}
     \Delta E = \Delta E_{\text{bulk}} +  \Delta E_{\text{edge}},
\end{equation}
where 
\begin{subequations}
\begin{equation}\label{eq_bulk_sigma}
\Delta E_{\rm{bulk}}  =  \sum_{j=1}^{L^2 - l} \Delta E_j = 4 \Delta \sigma_{\rm{bulk}} \sqrt{n},
\end{equation}
\begin{equation}\label{eq_edge_sigma}
\Delta E_{\text{edge}} =  \sum_{j=L^2-l+1}^{L^2} \Delta E_j = 4 \Delta \sigma_{\text{edge}} \sqrt{n}.
\end{equation}
\end{subequations}
Thus as a function of $\kappa$, i.e. the coupling between the spin and the fermion, $\Delta \sigma$ can be separately decomposed into its bulk and edge contributions where 
\begin{equation}
    \Delta \sigma = \Delta \sigma_{\text{bulk}} + \Delta \sigma_{\text{edge}}.
\end{equation}
While in the main text, the variation of $\Delta \sigma$ was shown as a function of $\kappa$, in Fig.~\ref{fig:edge_bulk_surface_tension}(a) we show the individual contributions from both the bulk and edge states. Deep in the topological regime $\kappa \sim 1$, one finds that the contribution to $\Delta \sigma$ is dominantly from the edge states. This is reflected even in the $\Delta E_j$, where we show its variation as a function of the eigenvalue index $j$ for $\kappa = 1.2$, which shows that only $\sim l$ states near the Fermi energy have significant $\Delta E_j$ (see Fig.~\ref{fig:edge_bulk_surface_tension}(b)).   In contrast, when $\kappa=2.5$ such that the system is in a trivial phase (no edge states on the droplet boundary), the sole contribution to $\Delta \sigma$ is due to the bulk states. The behavior of $\Delta E_{j}$ is non-universal and fluctuates over all the bulk states in the trivial phase (see Fig.~\ref{fig:edge_bulk_surface_tension}(c)).

A comparison of the $\Delta \sigma_{\text{edge}}$ with the analytic estimate $\Delta {\mathcal{E}}_{\rm{edge}}$ from the last section (where we set $\hbar=1$) is shown in Fig.~\ref{fig:edge_bulk_surface_tension}(d), illustrating the topological origin of the interface energy in the regime $0<|\kappa|<2$ which is universal and applicable to any topological system with a boundary manifold. Any contribution to $\Delta \sigma$ in the trivial phase $|\kappa|>2$ is due to the bulk states, which is model dependent and non-universal as discussed in the section below.

{\subsection{Non-Universal Interface Energy Contribution in Trivial Insulators}}

We now discuss how the contribution to $\Delta \sigma_{\text{bulk}}$ is non-universal and model dependent using two different Hamiltonian systems.\\

\begin{figure}
\centering
\includegraphics[width=1.0\columnwidth]{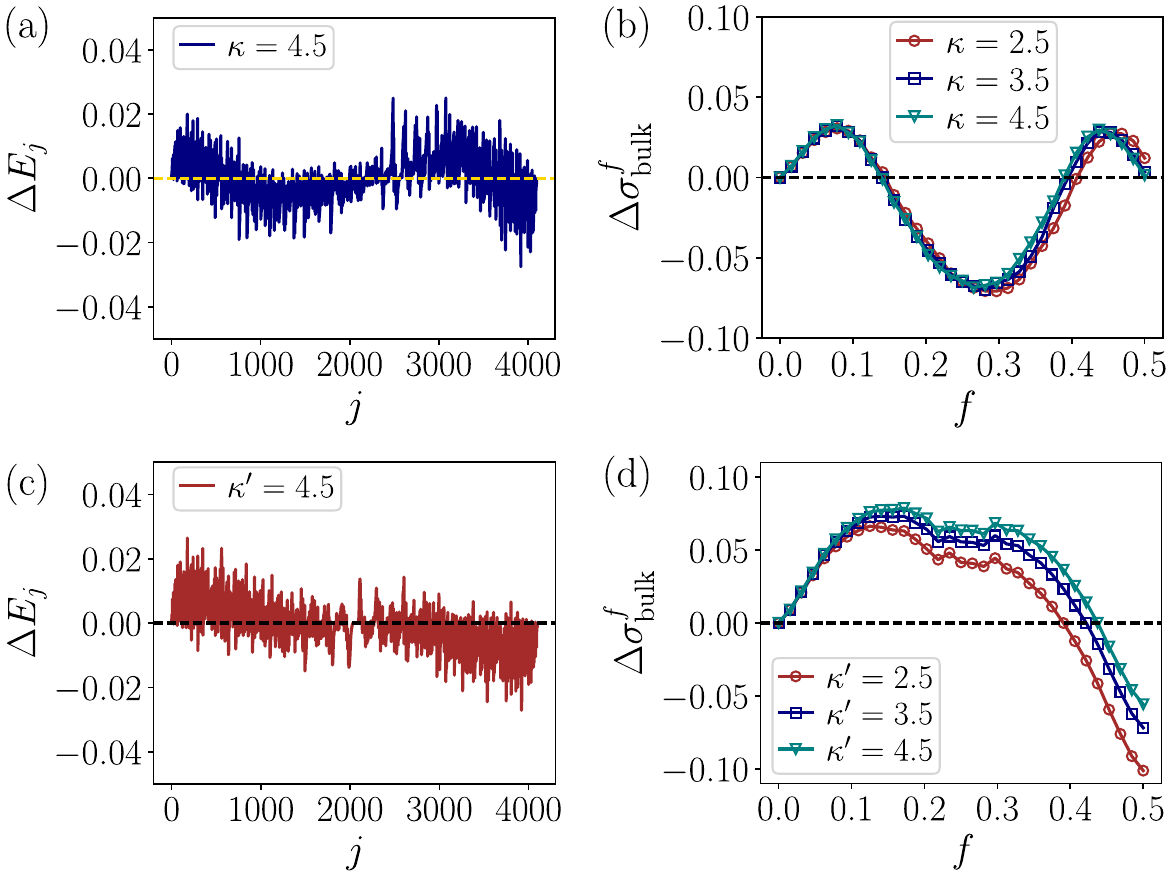}
\centering
\caption{
{\textbf{Correction to interfacial line energy for various filling fractions in trivial situation:}} (a) Difference of energy $\Delta E_{j}$ of $j$-th single-particle level (see Eq.~\eqref{eq_before_after}) after and before the quench in trivial situation ($\kappa=4.5$) when there is a square shaped droplet of size $n=l^{2}$ (where $l=16$) with $s_{i}=1$, while the rest of $s_{i}=-1$. (b) Correction to interfacial line energy $\Delta \sigma_{\rm{bulk}}^{f}$ as a function of filling fraction $f$ for various values of $\kappa$ in trivial situation of BHZ model. (c) $\Delta E_{j}$ as a function of $j$ in trivial insulator with $\kappa^{\prime}=4.5$ (see Hamiltonian in Eq.~\eqref{eq_trivialH1}) when there is a square shaped droplet of size $n=l^{2}$ (where $l=16$). (d) $\Delta \sigma_{\rm{bulk}}^{f}$ as a function of filling fraction $f$ for various values of $\kappa^{\prime}$ in trivial insulator . In all plots, number of unit cells is $L^{2}=4096$.}
\label{fig:supp_origin}
\end{figure}

\subsubsection{Trivial phases in BHZ model}

We first analyze the variation of the interfacial line energy correction in trivial situation ($|\kappa|>2$) of the model discussed in the main text, where the $\Delta E_j$ for $\kappa=2.5$ was shown in Fig.~\ref{fig:edge_bulk_surface_tension}(c). In Fig.~\ref{fig:supp_origin}(a) we show $\Delta E_j$ for $\kappa=4.5$, again reflecting the non-universal fluctuation of $\Delta E_j$ about zero. Furthermore, we find that both the value and the sign of $\Delta \sigma_{\text{bulk}}$ can be filling ($f$) dependent. In Fig.~\ref{fig:supp_origin}(b), we plot $\Delta \sigma_{\text{bulk}} ^{f}$ using
\begin{equation}
    \Delta E (f) = \sum_{j=1}^{j_{\rm{max}}} \Delta E_{j} = 4 (\Delta \sigma^f_{\text{bulk}}) \sqrt{n},
\end{equation}
where $j_{\rm{max}}=2 L^{2} f$ for different values of $\kappa=2.5,~3.5,~4.5$ all in the trivial phase of the model discussed in the main text. We find that $\Delta \sigma^f_{\text{bulk}}$ switches sign as $f$ is varied. While in this model all the bulk band gap closings are Dirac like - we investigate $\Delta \sigma^f_{\text{bulk}}$ in another model below where the bulk gap opening is quadratic.

\subsubsection{Quadratic trivial band insulator}

To demonstrate the non-universality of $\Delta \sigma_{\text{bulk}}$ in absence of any edge-mode, we consider another model represented by the Hamiltonian
\begin{align}\label{eq_trivialH1}
 H=  - \kappa^{\prime} \sum_{i} s_{i} {\bf{\Psi}}_{i}^{\dagger} \sigma_{z} ~{\bf{\Psi}}_{i} - \frac{1}{2}\sum_{\langle ij \rangle} ({\bf{\Psi}}_{i}^{\dagger} \sigma_{z}~{\bf{\Psi}}_{j} + {\rm{h.c.}}) \nonumber \\
 -J\sum_{\langle ij \rangle} s_{i} s_{j} -h \sum_{i} s_{i},
 \end{align}
 where $\pm \kappa^{\prime} s_{i}$ is the staggered mass of $A$ and $B$ sites of $i$-th unit cell and the hopping is restricted only between $A$-$A$ and $B$-$B$ sites of the nearest neighboring unit cells. For uniform $s_{i}=1$, we assume $M^{\prime} \equiv \kappa^{\prime} $ and the Hamiltonian can be recast in momentum space for $k_{x}, k_{y} \in (-\pi, \pi]$ as
 \begin{equation}
H_{k_{x}, k_{y}}= (-M^{\prime} - \cos(k_{x}) -\cos(k_{y})) \sigma_{z}.
 \end{equation}
Thus, the bulk energy-gap is zero for $|M^{\prime}| \leq 2$ where the system behaves as a metal, while for $|M^{\prime}|>2$ the bulk energy-gap is finite and the system is a trivial insulator. Just like the model presented in the main text the spectrum has a $M^{\prime} \to -M^{\prime}$ symmetry but at $M^{\prime}= \pm 2$ the metallic phase gives way to an insulator at half-filling via quadratically dispersing band. Importantly, there is no topological phase in this model. We now consider the same quench protocol as performed for BHZ model discussed in the main text. For a trivial insulator ($|\kappa^{\prime}| \gg 2$), we find that $\Delta E_{j}$ for all the bulk states shows non-universal fluctuation about zero  (see Fig.~\ref{fig:supp_origin}(c)). Similar to the trivial phase of BHZ model, here also $\Delta \sigma_{\rm{bulk}}^{f}$ shows non-universal dependence on filling $f$ and $\Delta \sigma_{\rm{bulk}}^{f}<0$ at half-filling (see Fig.~\ref{fig:supp_origin}(d)).

Therefore, the correction to the interfacial line energy $\Delta \sigma_{\rm{bulk}}$ in trivial phase at half-filling (i.e. ground state) can be either positive or negative which is decided by the random contributions from individual single-particle energy-levels, rendering the behavior of the interfacial line energy in trivial situation non-universal.

\section{Role of Symmetries and Topological Interface Energy}\label{app_symmetry}
Here we demonstrate the role of breaking spin-fermion flip symmetry and particle-hole symmetry in affecting the interfacial line energy in topological phases.

{{
{\subsection{Role of breaking spin-fermion flip symmetry (class D model)}}}}

\begin{figure}
    \centering
\includegraphics[width=1.0\linewidth]{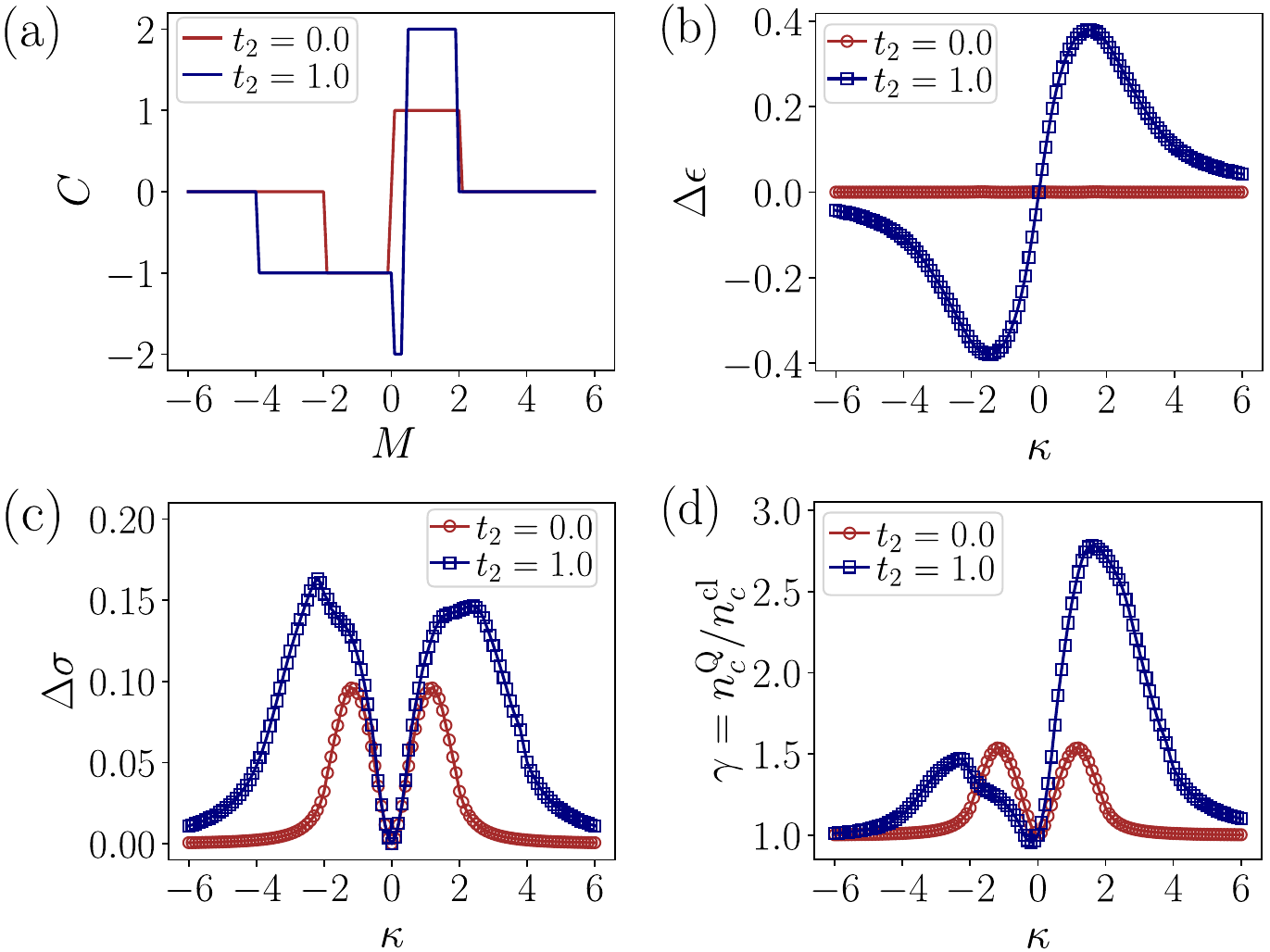}
\caption{{{{\textbf{Effect of hopping between next nearest neighbors:}}{ (a) Chern number $C$ as a function of mass parameter $M$, (b) change in energy density $\Delta \epsilon$ as a function of $\kappa$, (c) additional interfacial line energy $\Delta \sigma$ as a function of $\kappa$, (d) enhancement ratio $\gamma=n_{c}^{\rm{Q}}/n_{c}^{\rm{cl}}$ of critical cluster size
as a function of $\kappa$ when the hopping between next nearest neighboring unit cells is absent ($t_{2}=0.0$) and present ($t_{2}=1.0$). In all plots, $J=0.2$, $h=1.0$, $T=0$, $L^{2}=4096$.}}}}\label{fig_nnn}
\end{figure}

{{In order to break the spin-fermion flip symmetry (while retaining particle-hole symmetry), we introduce the next-nearest neighbor hopping $t_2$ such that the complete spin-fermion hopping Hamiltonian becomes
\begin{align}\label{eq_hnnn}
H &=- \kappa \sum_{i} s_{i} {\bf{\Psi}}_{i}^{\dagger} \sigma_{z} ~{\bf{\Psi}}_{i} - \sum_{\langle ij \rangle} ({\bf{\Psi}}_{i}^{\dagger} {\large{\eta}}_{ij}~{\bf{\Psi}}_{j} + {\rm{h.c.}}) \nonumber \\
&- t_{2} \sum_{\langle \langle ij \rangle \rangle} ({\bf{\Psi}}_{i}^{\dagger} {\large{\eta}}^{\prime \prime}_{ij}~{\bf{\Psi}}_{j} + {\rm{h.c.}})-J\sum_{\langle ij \rangle} s_{i} s_{j} -h \sum_{i} s_{i},
\end{align}
where $t_{2}$ is the strength of the hopping between the next nearest neighboring unit cells and $\eta^{\prime \prime}_{ij}=\frac{1}{2}(\sigma_{z}+ i \cos(\theta_{ij}) \sigma_{x} + i \sin(\theta_{ij}) \sigma_{y})$, the angle $\theta_{ij}$ between the $x$-axis and the bond connecting the next nearest neighboring unit cells can assume the values $\theta_{ij}=\pi/4$ and $\theta_{ij}=3\pi/4$. For uniform $s_{i}$ throughout the system, we assume $M\equiv \kappa s_{i}$ and the Hamiltonian in momentum space for $k_{x}, k_{y} \in (-\pi,\pi]$ is written as
\begin{align}
H_{k_{x},k_{y}} &= d_{1} (k_{x}, k_{y})\sigma_{x} +  d_{2} (k_{x}, k_{y}) \sigma_{y} +  d_{3} (k_{x}, k_{y}) \sigma_{z}, \nonumber \\
d_{1}(k_{x},k_{y}) &= \sin(k_{x}) -\sqrt{2} t_{2} \cos(k_{x}) \sin(k_{y}), \nonumber \\
d_{2}(k_{x},k_{y}) &= \sin(k_{y}) -\sqrt{2} t_{2} \cos(k_{y}) \sin(k_{x}), \nonumber \\
d_{3}(k_{x},k_{y}) &= -M -\cos(k_{x})-\cos(k_{y})-2 t_{2} \cos(k_{x}) \cos(k_{y}).
\end{align}
The Hamiltonian satisfies particle-hole symmetry even when $t_{2} \neq 0$, which follows from
\begin{equation}
\sigma_{x} H^{*}_{k_{x},k_{y}} \sigma_{x}=-H_{-k_{x},-k_{y}}.
\end{equation}
For uniform $s_{i}$, the system can host topological phases having Chern numbers $C=\pm 2, \pm 1$ depending on the value of mass parameter $M$ when $t_{2} \neq 0$ (see Fig.~\ref{fig_nnn}(a)). Interestingly, the ground state energy of the fermionic system, which is given by
\begin{align}
E_{g} (M)  = \sum_{k_{x}, k_{y}} E_{k_{x},k_{y}} & =\frac{L^{2}}{(2\pi)^{2}} \int_{-\pi}^{\pi} dk_{x} \int_{-\pi}^{\pi} dk_{y} E_{k_{x},k_{y}} \nonumber \\
\text{where~} E_{k_{x},k_{y}} &= -\sqrt{d_{1}^{2}+ d_{2}^{2}+ d_{3}^{2}}~,
\end{align}
is not symmetric under $M \to -M$, i.e., $E_{g}(M) \neq E_{g}(-M)$ when $t_{2} \neq 0$. Thus, the sudden quench from $h<0$ to $h>0$ causes an additional correction to bulk energy density $\Delta \epsilon$ along with the correction to interfacial line energy $\Delta \sigma$ when $t_{2} \neq 0$, such that the change in the energy in the fermionic system for a formation of a square-shaped droplet of size $n$ is
\begin{equation}
    \Delta E = (\Delta \epsilon) n + 4 \sqrt{n} (\Delta \sigma).
\end{equation}
We show the variation of $\Delta \epsilon$ and $\Delta \sigma$ in Fig.~\ref{fig_nnn}(b,c), where we observe that unlike the situation with $t_{2}=0$, $\Delta \epsilon \neq 0$ for $\kappa \neq 0$ when $t_{2} \neq 0$. Furthermore, in the topological situation $0<|\kappa|<2$, $\Delta \sigma$ for $t_{2} \neq 0$ is greater than that for $t_{2}=0$. This can be understood as follows: when $t_{2} \neq 0$ and $0<\kappa<2$, a droplet of $s_{i}=1$ corresponds to Chern number $C=2$, while its background with $s_{i}=-1$ has $C=-1$. Thus, the difference in Chern numbers ($|\Delta C|=3$) between the droplet and its background for $t_{2} \neq 0$ is greater than $|\Delta C|=2$ for $t_{2}=0$. An increased $|\Delta C|$ in topological situation causes greater $\Delta \sigma$ since the number of edge modes increases, increasing the effective $\Delta {\mathcal E}_{\text{edge}}$.  This results in a higher $\Delta \sigma$ for $t_{2} \neq 0$ compared to that for $t_{2}=0$. For $-2<\kappa<0$, $\Delta \sigma$ shows similar behavior where a droplet of $s_{i}=1$ has $C=-1$ and its background with $s_{i}=-1$ has $C=2$. In presence of non-zero $\Delta \epsilon$ and $\Delta \sigma$, the critical cluster size for $t_{2} \neq 0$ modifies to
\begin{equation}
    n_{c}^{\rm{Q}} = \frac{\pi (\sigma_{\rm{cl}}+ \Delta \sigma)^{2}}{(2 |h| -\Delta \epsilon)^{2}},
\end{equation}
resulting in an enhancement ratio
\begin{equation}
    \gamma=\frac{n_{c}^{\rm{Q}}}{n_{c}^{\rm{cl}}}=\frac{(1+(\Delta \sigma/\sigma_{\rm{cl}}))^{2}}{(1-(\Delta \epsilon/(2|h|)))^{2}}.
\end{equation}
We find that in topological situation, $\gamma$ can be greater than $1$ for a suitable choice of $J$ and $h$, indicating an enhancement of critical cluster size (see Fig.~\ref{fig_nnn}(d)).
}}

\begin{figure}
    \centering
\includegraphics[width=1.0\linewidth]{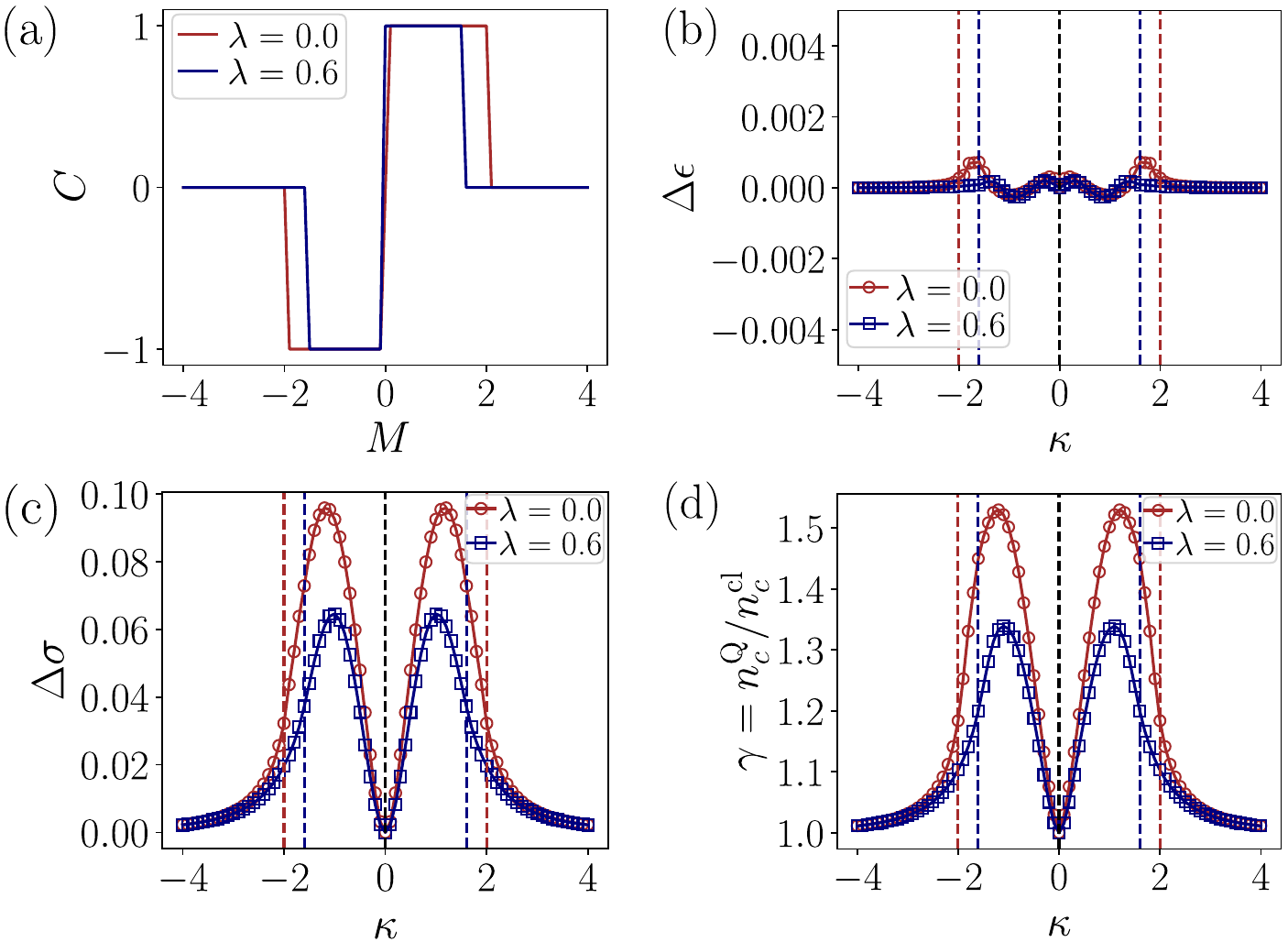}
\caption{{{\textbf{Effect of particle-hole symmetry breaking hopping term:}} {(a) Chern number $C$ as a function of mass parameter $M$, (b) change in energy density $\Delta \epsilon$ as a function of $\kappa$, (c) additional interfacial line energy $\Delta \sigma$ as a function of $\kappa$, (d) enhancement ratio $\gamma=n_{c}^{\rm{Q}}/n_{c}^{\rm{cl}}$ of critical cluster size
as a function of $\kappa$ when particle-hole symmetry breaking hopping term is absent ($\lambda=0.0$) and present ($\lambda=0.6$). Dotted lines denote the critical points $\kappa=0, \pm 2$ when $\lambda=0$ and $\kappa=0, \pm 1.6$ when $\lambda=0.6$. In all plots, $J=0.2$, $T=0$, $L^{2}=4096$.}}}\label{fig_ph_broken}
\end{figure}

\begin{figure*}
\centering
\includegraphics[width=1.0\linewidth]{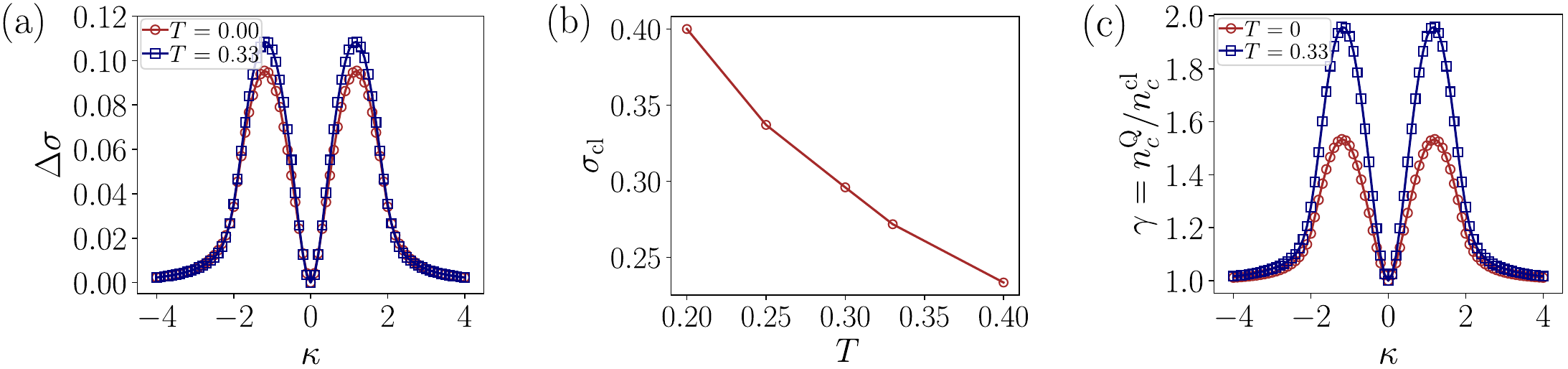}
\centering
\caption{{\textbf{Interfacial line energy and critical cluster-size at finite temperature:}} (a) Quantum correction to the interfacial line energy $\Delta \sigma$ as a function of $\kappa$ at temperatures $T=0$ and $T=1/\beta=0.33$, (b) $\sigma_{\rm{cl}}$ of classical Ising model ($\kappa=0$) as a function of $T$, (c) $\gamma=n_{c}^{\rm{Q}}/n_{c}^{\rm{cl}}=(1+(\Delta \sigma/\sigma_{\rm{cl}}))^{2}$ as a function of $\kappa$ at temperatures $T=0$ and $T=1/\beta=0.33$, where $n_{c}^{\rm{cl}}$ is the critical cluster-size for classical Ising model ($\kappa=0$). For all the plots, number of unit cells is $L^{2}=4096$. The parameters chosen are: $J=0.2$ (where $T_{c} \approx 0.45$) and $h=0.02$. Here, $\sigma_{\rm{cl}} \sim 2J$ at $T=0$. At finite $T$ in (b), $n_{c}^{\rm{cl}}$ is evaluated using Monte Carlo simulation for classical Ising model ($\kappa=0$) via Eq.~\eqref{eq_nu_clT} and $\sigma_{\rm{cl}}$ is then obtained from $n_{c}^{\rm{cl}}$ using Eq.~\eqref{eq_scl}. In (c), $n_{c}^{\rm{cl}}=145$ for $T=1/\beta=0.33$.}
\label{fig_finiteT}
\end{figure*}

{\subsection{ Role of breaking particle-hole symmetry (class A  model)}}

{{To explore the effect of the hopping term that breaks particle-hole symmetry in BHZ model, we now consider the Hamiltonian
\begin{align}\label{eq_ph}
H &=- \kappa \sum_{i} s_{i} {\bf{\Psi}}_{i}^{\dagger} \sigma_{z} ~{\bf{\Psi}}_{i} - \sum_{\langle ij \rangle} ({\bf{\Psi}}_{i}^{\dagger} {\large{\eta}}_{ij}~{\bf{\Psi}}_{j} + {\rm{h.c.}}) \nonumber \\
&- \lambda \sum_{i} {\bf{\Psi}}_{i}^{\dagger} (\sigma_{x}+\sigma_{y})~{\bf{\Psi}}_{i}-J\sum_{\langle ij \rangle} s_{i} s_{j} -h \sum_{i} s_{i},
\end{align}
where $\lambda$ is the hopping parameter between $A$ and $B$ sites within the same unit cell which breaks the particle-hole symmetry. For uniform $s_{i}$ throughout the system, we assume $M\equiv \kappa s_{i}$ and the Hamiltonian in momentum space for $k_{x}, k_{y} \in (-\pi,\pi]$ turns out to be
\begin{align}
H_{k_{x},k_{y}} = d^{\prime}_{1} (k_{x}, k_{y})\sigma_{x} &+  d^{\prime}_{2} (k_{x}, k_{y}) \sigma_{y} +  d^{\prime}_{3} (k_{x}, k_{y}) \sigma_{z}, \nonumber \\
d^{\prime}_{1}(k_{x},k_{y}) &= (\sin(k_{x})-\lambda), \nonumber \\
d^{\prime}_{2}(k_{x},k_{y}) &= (\sin(k_{y})-\lambda), \nonumber \\
d^{\prime}_{3}(k_{x},k_{y}) &= (-M -\cos(k_{x})-\cos(k_{y})).
\end{align}
For $\lambda \neq 0$, the particle-hole symmetry of the Hamiltonian is broken, i.e., $\sigma_{x} H^{*}_{k_{x},k_{y}} \sigma_{x} \neq -H_{-k_{x},-k_{y}}$. When $\lambda \neq 0$, the system with uniform $s_{i}$ hosts topological phases with $C=\pm 1$ for $0<|M|<2 \sqrt{1- \lambda^{2}}$ and trivial phase ($C=0$) for $|M|>2 \sqrt{1- \lambda^{2}}$ (see Fig.~\ref{fig_ph_broken}(a)). Even in absence of particle-hole symmetry ($\lambda \neq 0$), the ground state energy of the fermionic system with uniform $s_{i}$ remains symmetric under $M \to - M$, i.e., $E_{g}(M)=E_{g}(-M)$. Thus, the sudden quench from $h<0$ to $h>0$ causes no additional correction to bulk energy density ($\Delta \epsilon \to 0$),  while there is a substantial correction to the interfacial line energy $\Delta \sigma>0$ in topological situation (see Fig.~\ref{fig_ph_broken}(b,c)). Therefore, the enhancement ratio $\gamma=n_{c}^{\rm{Q}}/n_{c}^{\rm{cl}}=(1+(\Delta \sigma/\sigma_{\rm{cl}}))^{2}$ for $\lambda \neq 0$ shows similar behavior as for $\lambda=0$ (see Fig.~\ref{fig_ph_broken}(d)) and establishes the increase in critical cluster-size even when particle-hole symmetry is absent.}}

\section{Role of thermal fluctuations}\label{app_thermal}

Now we discuss the effects of thermal fluctuations on the interfacial line energy and critical droplet size both for the fermions and for the spin degrees of freedom.

\subsection{Fermionic fluctuations}

In order to distill the role of thermal fluctuation on fermions we freeze the spin configurations before and after quench to where all $s_i=-1$ and where a region of size $n=l^2$ has $s_i=1$. However fermions are now assumed to equilibrate to a finite temperature $T$. It is useful to note we work with $T< T_c$ where classical transition temperature $T_{c} \approx 2.269 J$~\cite{onsager_physrev_1944} (where Boltzmann constant is set to unity).

The correction to the interfacial line energy at a finite temperature $\Delta \sigma (T)$ is again computed from $\Delta E (T) = 4 \Delta \sigma (T) \sqrt{n}$, where the change in energy $\Delta E(T)$ for the fermionic problem at temperature $T=1/\beta$ (where Boltzmann constant is set to unity) is given by
\begin{equation}
\Delta E(T)= \frac{{\rm{Tr}}(e^{-\beta H_{\rm{final}}}~H_{\rm{final}})}{{\rm{Tr}}(e^{-\beta H_{\rm{final}}})}-\frac{{\rm{Tr}}(e^{-\beta H_{\rm{initial}}}~H_{\rm{initial}})}{{\rm{Tr}}(e^{-\beta H_{\rm{initial}}})}.
\end{equation}
Here, $H_{\rm{initial}}$ is the Hamiltonian for fermionic problem before the quench where all $s_{i}=-1$ and $H_{\rm{final}}$ is the Hamiltonian after the quench where there is a square-shaped cluster of size $n=l^{2}$ having $s_{i}=1$ within the region having $s_{i}=-1$. We find that at a finite temperature $T$, $\Delta \sigma(T)$ does not change significantly from its zero temperature ($T=0$) value in both the topological ($0<|\kappa|<2$) and trivial ($|\kappa|>2$) situations, as evident from Fig.~\ref{fig_finiteT}(a). This is expected since the thermal fluctuations at $T<T_c$ are perturbative compared to the fermionic bandwidth.

\begin{figure*}
    \centering
\includegraphics[width=1.0\linewidth]{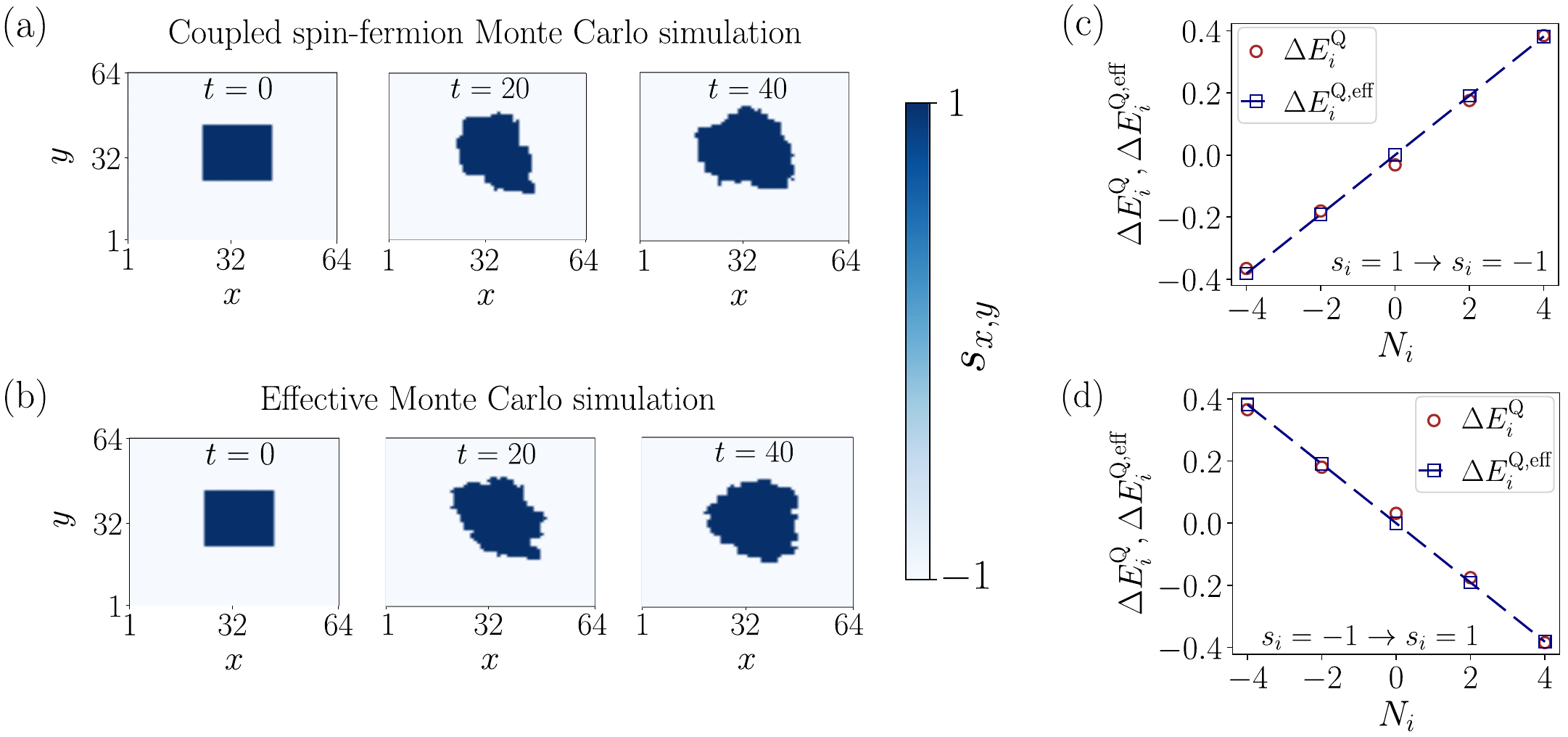}
\caption{{\textbf{Effective Monte Carlo simulation:}} (a,b)  Local magnetization $ s_{x,y}$ for unit cell with position $(x,y)$ at times $t=0,20,40$ (in units of Monte Carlo steps per site) using (a) coupled spin-fermion Monte Carlo simulation (Eq.~\eqref{eq_diff_qme} and Eq.~\eqref{eq_qm_mc}) and (b) effective Monte Carlo simulation (Eq.~\eqref{eq_qm_Ei} and Eq.~\eqref{eq_eff_mc}). Here, the initial cluster-size with up-spins is $n=441$ ($n>n_{c}^{\rm{Q}}$). The parameters chosen are: $\kappa=1.2$, $J=0.2$, $T=1/\beta=0.33$ (where $T_{c} \approx 0.45$), $h=0.02$. (c,d) Change in ground state energy $\Delta E_{i}^{\rm{Q}}$ (see Eq.~\eqref{eq_diff_qme}) and its comparison with $\Delta E_{i}^{\rm{Q, eff}}$ (see Eq.~\eqref{eq_eff_mc}) for fermionic system when $i$-th spin is flipped from (c) $s_{i}=1$ to $s_{i}=-1$, (d) $s_{i}=-1$ to $s_{i}=1$. Here, $N_{i}=\sum_{j \in {\mathcal{N}}_{i}} s_{j}$ where $j$ runs over all the four nearest neighbors (${\mathcal{N}}_{i}$) of $i$-th spin. We find that $\Delta E_{i}^{\rm{Q}} \approx \Delta E_{i}^{\rm{Q, eff}}$. Here, we have chosen $\kappa=1.2$ and number of unit cells in square lattice $L^{2}=4096$.}\label{fig_mc_check}
\end{figure*}

\subsection{Entropic contributions to spin fluctuations}
In order to estimate the role of thermal fluctuations on classical $\sigma_{\rm{cl}}$ which is $\sim 2J$ at $T=0$, we perform the Monte Carlo (MC) simulations. Here, a finite-sized cluster is subjected to the non-equilibrium situation (after a sudden quench of $h$) and the critical cluster-size $n_{c}^{\rm{cl}}$ at $\kappa=0$ is evaluated via 
\begin{equation}\label{eq_nu_clT}
 \nu(n_{c}^{\rm{cl}}) = \frac{P_+ (n_{c}^{\rm{cl}}) - P_-(n_{c}^{\rm{cl}})}{P_+ (n_{c}^{\rm{cl}}) + P_-(n_{c}^{\rm{cl}})} = 0  
\end{equation}
where $P_{+}(n)$ and $P_{-}(n)$ are the number of MC steps in which size $n$ of a cluster increases to $(n+1)$ and decreases to $(n-1)$ respectively~\cite{katsuno_pre_2011}. Here, for the cluster-size $n<n_{c}^{\rm{cl}}$, the tendency of shrinkage of the cluster is greater than that of growth, implying $P_{+}<P_{-}$ and $\nu<0$,  while for $n>n_{c}^{\rm{cl}}$, the tendency of growth of the cluster is greater than that of shrinkage, implying $P_{+}>P_{-}$ and $\nu>0$. Therefore, $\nu=0$ for $n=n_{c}^{\rm{cl}}$. 

From the critical cluster size $n_{c}^{\rm{cl}}$, we then estimate $\sigma_{\rm{cl}} (T)$
using
\begin{equation}\label{eq_scl}
\sigma_{\rm{cl}} (T) =2 |h| \sqrt{\frac{n_{c}^{\rm{cl}} }{\pi}},
\end{equation}
since any correction to $\sigma_{\rm{cl}} (T)$ for $\kappa=0$ is only due to thermal spin fluctuations and has no fermionic contribution. As shown in Fig.~\ref{fig_finiteT}(b), finite temperature $T$ reduces $\sigma_{\rm{cl}}$ due to entropic contributions coming from the spin fluctuations in the nucleating region~\cite{kevin_pre_2005}. Given the reduction in the effective $\sigma_{\rm{cl}}$ with $T$, this illustrates that the thermal fluctuations on the  boundary of the nucleating region effectively reduces the stiffness of the domain wall.

Interestingly, the effect of thermal fluctuations on fermionic and spin together lead to an increase in the relative enhancement $\gamma=n_{c}^{\rm{Q}}/n_{c}^{\rm{cl}}=(1+(\Delta \sigma/\sigma_{\rm{cl}}))^{2}$ due to an effective reduction of the $\sigma_{\rm{cl}}$. This enhancement of $\gamma$ at finite $T$ is more dominant in the topological situation ($0<|\kappa|<2$) compared to the trivial situation ($|\kappa|>2$), given  the fermionic contribution $\Delta \sigma$ is significant in the topological regime (see Fig.~\ref{fig_finiteT}(c)).

\begin{figure*}
    \centering
\includegraphics[width=1.0\linewidth]{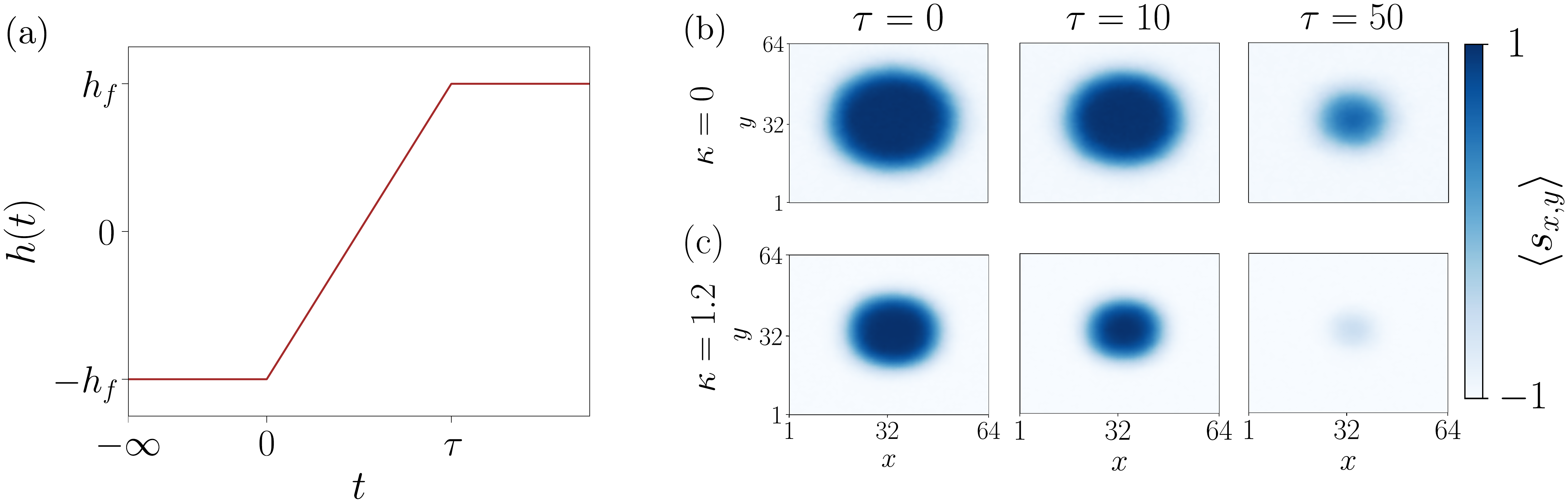}
    \caption{
    {\textbf{Dynamics of average local magnetization under slow quench:}}
    (a) Linear quench protocol of $h(t)$ as a function of $t$ (see Eq.~\eqref{eq_slow_qn}),
    (b,c) Average local magnetization $\langle s_{x,y} \rangle$ for the unit cell with coordinate $(x,y)$ for a cluster of up-spins having initial size $n=441$ at time $t=100$ (in the units of Monte Carlo steps per site) for different values of $\tau$ when (b) $\kappa=0$ and (c) $\kappa=1.2$. In all the plots, the parameters chosen are: $J=0.2$, $T=1/\beta=0.33$ (where $T_{c} \approx 0.45$), $h_{f}=0.02$. The number of unit cells in square lattice is $L^{2}=4096$ and number of simulations considered is $1000$. Cluster size at fixed $t=100$ reduces with increasing $\tau$.}
    \label{fig_slow}
\end{figure*}

\section{Effective Monte Carlo simulation and validity}\label{app_monte_carlo}

In order to simplify the exact MC method, we here show that we can also calculate $\Delta E_{i}$ approximately for a spin-flip ($s_i \to -s_i$) when $\kappa \neq 0$ from the quantum correction to the interfacial line energy $\Delta \sigma$ as:
\begin{equation}\label{eq_qm_Ei}
\Delta E_{i}=2Js_{i} \sum_{j \in {\mathcal{N}}_{i}} s_{j}+2h s_{i} + \Delta E^{\rm{Q,eff}}_{i},
\end{equation}
where 
\begin{equation}\label{eq_eff_mc}
\Delta E^{\rm{Q,eff}}_{i} = (\Delta \sigma) s_{i} \sum_{j \in {\mathcal{N}}_{i}} s_{j},
\end{equation}
where $\Delta \sigma$ is evaluated at $T=0$. This effectively replaces the fermion diagonalization step with an effective correction from the fermions. In Fig.~\ref{fig_mc_check}(a,b), we show the exact and the effective MC dynamics with the same initial random seed and same initial configurations for $\kappa=1.2$. We next estimate the validity of this step by comparing $\Delta E_{i}^{\rm{Q}}$ from each Monte Carlo dynamics (see Eq.~\eqref{eq_diff_qme}) with effective $\Delta E_{i}^{\rm{Q,eff}}$. The energy correction can be estimated for the two processes $s_i =1 \rightarrow -1$ and for $s_i = -1 \rightarrow 1$  for different configurations of neighborhood spins characterized by $N_{i}= \sum_{j \in {\mathcal{N}}_{i}} s_{j}$ that can take values $-4, -2, 0, 2, 4$. In Fig.~\ref{fig_mc_check}(c,d), we show the variation of $\Delta E^{\rm{Q,eff}}_{i}$ and $\Delta E_{i}^{\rm{Q}}$ for both the processes for different values of $N_{i}$ at $\kappa = 1.2$. The values are essentially indistinguishable, establishing the validity and utility of the effective MC procedure. Thus, while exact droplet growth can have minor fluctuations between exact MC dynamics and effective MC dynamics, the effective MC method is particularly apt in evaluating the coarse-grained description of the system.

\section{Role of Slow Quench}\label{slow}

We now analyze the growth and shrinkage of clusters under slow quench protocol of the external magnetic field $h(t)$ where we consider
\begin{equation}\label{eq_slow_qn}
h(t)=\begin{cases}
-h_{f}+\frac{2h_{f} t}{\tau}, \text{~~~~for $0<t<\tau$,}\\
~~~h_{f}, \text{~~~~~~~~~~~~~for $t \geq \tau$.}
\end{cases}
\end{equation}
Thus, $\tau$ is the timescale of the quench (see Fig.~\ref{fig_slow}(a)) over which $h(t)$ changes from $-h_{f}$ to $h_{f}$ (where we assume $h_{f}>0$) and the limit $\tau=0$ corresponds to the sudden quench considered in the main text. If $t$ and $\tau$ are chosen in the units of Monte Carlo steps per site, then we find that the size of a cluster having up-spins at a fixed time $t>\tau$ decreases with the increase of $\tau$ for both classical ($\kappa=0$) and topological ($0<|\kappa|<2$) situations, as can be seen from average local magnetization $\langle s_{x,y} \rangle$ in Fig.~\ref{fig_slow}(b,c). This observation can be explained as follows: in this slow quench protocol, when $t<\tau/2$, $h(t)<0$  and thus a cluster having up-spins is not in the global free-energy minima and reduction of its size (i.e. growth of its larger background region) is more favored. Later, when $h(t)>0$ (for $t>\tau/2$), the cluster having up-spins reaches the global free-energy minima and starts showing growth (shrinkage) determined by usual nucleation physics. However, due to the reduction of its size when $t<\tau/2$, the size of a cluster at a fixed time $t>\tau$ becomes smaller compared to that for sudden quench situation ($\tau=0$).
Thus, if a cluster grows under a sudden quench, it can be made to shrink when the quench protocol is changed to a slow quench performed over a time $\tau$, since the nucleation phenomenon is decided by the adiabatic limit where the instantaneous free-energy landscape determines the nucleation process. This leads to an increase in the effective critical cluster size with increasing $\tau$ in the slow quench protocol. Thus, the timescale $\tau$ can effectively tune the critical cluster size and influence the growth (shrinkage) of clusters.

\bibliography{reference.bib}

\end{document}